\documentclass[iop,twocolappendix,numberedappendix]{emulateapj}
\usepackage{natbib}
\usepackage[usenames,dvipsnames]{color}
\usepackage{gensymb}
\tabletypesize{\tiny}
\usepackage[Symbol]{upgreek}

\usepackage{hyperref}
\hypersetup{colorlinks=true}
\extrafloats{100}

\newcommand{\twopiskytm}{2$\uppi$Sky\textsuperscript{TM}}
\newcommand{\owtm}{ObservatoryWatch\textsuperscript{TM}}

\newcommand{\bjdtdb}{\ensuremath{\rm {BJD_{TDB}}}}
\newcommand{\feh}{\ensuremath{\left[{\rm Fe}/{\rm H}\right]}}
\newcommand{\teff}{\ensuremath{T_{\rm eff}}}

\newcommand{\msun}{\ensuremath{\,M_\Sun}}
\newcommand{\rsun}{\ensuremath{\,R_\Sun}}
\newcommand{\lsun}{\ensuremath{\,L_\Sun}}
\newcommand{\mj}{\ensuremath{\,M_{\rm J}}}
\newcommand{\rj}{\ensuremath{\,R_{\rm J}}}
\newcommand{\fave}{\langle F \rangle}
\newcommand{\fluxcgs}{10$^9$ erg s$^{-1}$ cm$^{-2}$}

\newcommand{\ra}[4]{#1\textsuperscript{h}#2\textsuperscript{m}#3\fs#4}
\newcommand{\de}[4]{#1\degr#2\arcmin#3\farcs#4}

\slugcomment{Accepted for publication in the Publications of the Astronomical Society of the Pacific, 21 July 2017}

\shorttitle{Solaris Global Telescope Network}
\shortauthors{Koz\l owski et al.}

\begin{document}

\title{Project Solaris -- a Global Network of Autonomous Observatories -- Design,  Commissioning and First Science Results}

\author{S.K. Koz\l owski\altaffilmark{1,2}, P.W. Sybilski\altaffilmark{1,3}, M. Konacki\altaffilmark{1,2,3}, R.K. Paw\l aszek\altaffilmark{1,3}, M. Ratajczak \altaffilmark{4}, ~K.G.~He\l miniak\altaffilmark{1} and M. Litwicki\altaffilmark{1,2}}
\email{stan@ncac.torun.pl}

\altaffiltext{1}{Nicolaus Copernicus Astronomical Center, Bartycka 18, 00-716 Warsaw, Poland}
\altaffiltext{2}{Cilium Engineering Sp. z o.o., \L okietka 5, 87-100 Toru\'n, Poland}
\altaffiltext{3}{Sybilla technologies Sp. z o.o., Toru\'nska 59, 85-023 Bydgoszcz, Poland}
\altaffiltext{4}{Astronomical Institute, Univeristy of Wroc\l aw, Kopernika 11, 51-622 Wroc\l aw, Poland}

\begin{abstract}
We present the design and commissioning of Project Solaris, a global network of autonomous observatories. Solaris is a Polish scientific undertaking aimed at the detection and characterization of circumbinary exoplanets and eclipsing binary stars. To accomplish this task, a network of four fully autonomous observatories has beed deployed in the Southern Hemisphere: Solaris-1 and Solaris-2 in the South African Astronomical Observatory in South Africa, Solaris-3 in Siding Spring Observatory in Australia and Solaris-4 in Complejo Astronomico El Leoncito in Argentina. The four stations are nearly identical and are equipped with 0.5-m Ritchey-Cr\'{e}tien (f/15) or Cassegrain (f/9, Solaris-3) optics and high-grade 2K x 2K  CCD cameras with Johnson and Sloan filter sets. We present the design and implementation of low-level security, data logging and notification systems, weather monitoring components, all-sky vision system, surveillance system and distributed temperature and humidity sensors. We describe dedicated grounding and lighting protection system design and robust fiber data transfer interfaces in electrically demanding conditions. We discuss the outcomes of our design as well as the resulting software engineering requirements. We describe our systems engineering approach to achieve the required level of autonomy, the architecture of the custom high-level industry-grade software that has been designed and implemented specifically for the use of the network. We present the actual status of the project and first photometric results. These include data and models of already studied systems for benchmarking purposes (Wasp-4b, Wasp-64b and Wasp-98b transits, PG1336-016, an eclipsing binary with a pulsator) as well J024946-3825.6, an interesting low-mass binary system, for which a complete model is provided for the first time. 
\\
\\
\footnotesize
Accepted for publication in Publications od the Astronomical Society of the Pacific on July 21st 2017. 
\normalsize
\end{abstract}



\section{Introduction}
\label{sec:Introduction}

Various exoplanet detection methods exist that are sensitive to different star-planet configurations. In the past, exoplanets orbiting both components of a binary system (circumbinary planets) have been searched for using, for example, radial velocities (RVs), including the iodine cell technique applied to double-lined spectroscopic binaries \citep{Konacki2005,Konacki2009}, direct imaging \citep{Currie2014,Kraus2014,Bonavita2016}, or microlensing \citep{Bennett2016}. The first transiting circumbinary planet has been discovered by Doyle et al. (2011) with several more to follow since then, including the popular Kepler-47 system \citep{Orosz2012} or a quadruple star PH-1 \citep{Schwamb2013}.

The main goal of Project Solaris is to conduct an observing campaign in the search for circumbinary planets using a dedicated network of telescopes that can supply high cadence and high precision photometry data for (a) the utilization of the eclipse timing method and for  (b) characterization of eclipsing binary stars. The secondary goal of the Project is to increase the competence in the design and construction of robotic telescopes, associated hardware components and software. The name \textit{Solaris} is a tribute to Stanis\l aw Lem's identically titled novel published in 1961. In his book the famous Polish writer describes a fictional planet that exists in a binary system. To date, 25 planetary systems (31 planets, 5 multi planetary systems) have been detected strictly using timing methods (exoplanet.eu, June 2017), some of which are not widely accepted by the community due to low credibility of the results.  Possibly, more exoplanets will be discovered using transit timing variation in the Kepler data \citep{Borucki2010,Koch2010}.  

The timing technique with respect to exoplanets dates back to 1992, when the first exoplanet was discovered: PSR~1257~12b \citep{Wolszczan1992, Konacki2003}. Variations in the otherwise precisely periodic pulse signal have been identified as the footprint of the presence of an additional body in the system. The motion of the pulsar around the common center of mass and the finite speed of light cause the pulses appear to the observer being too early or too late, which is known as the light-time effect \citep{Sterken2005a,Sterken2005b}. 
In case of eclipsing binary stars, the regular eclipses act as the carrier signal instead of pulsar pulses. An additional body in the system will cause variations in timing of eclipses. The resulting light-time orbit can be derived from long-span photometric data.

The first Automated Photoelectric Telescopes have been built in the mid 1960's. Since then more and more telescopes operate without direct human supervision \citep{Castro2010}, sometimes during extended periods of time. Instruments attached to telescopes tend to be more complex and need to fulfill the most demanding requirements of astrophysicists in terms of efficiency, speed, precision and stability. In many cases manual operation of modern telescope systems is not even possible. This applies not only to the largest instruments but also to distributed networks of smaller telescopes, such as Project Solaris, that operates a global network of autonomous telescopes. In Sec \ref{sec:NetworkDescription} we describe the global network, locations of the observatories. In Sec. \ref{sec:HardwareComponents} we describe the hardware components of the individual sites. Section \ref{sec:SoftwareArchitecture} describes the software that is used to control, manage and operate the entire network. In Sec. \ref{sec:Operation} we focus on the operation of the network to date including major problems encountered during the installation and operation phases. We present scientific commissioning results in Sec. \ref{sec:FirstResults} and we summarize in Sec. \ref{sec:Summary}. 

\subsection{Autonomous Observatories and Existing Telescope Networks}
\label{ssec:ExistingNetworks}

Types of observatories based on their operation mode have been defined by \cite{Gelderman2001}. Observatories can be remote, unmanned, robotic and fully autonomous. This nomenclature refers mainly to the way observations are executed, i.e. how advanced are the scheduling algorithms. From the system's engineering point of view, however, the proposed classification is not complete. This becomes more evident when the the telescope is treated as a robot with two (usually) degrees of freedom that operates in a controlled environment. A detailed elaboration on the nomenclature is provided in Sec. \ref{sec:SoftwareArchitecture}. Two or more observatories located in different sites that operate within the same framework or are governed by the same institution constitute a network of observatories. In 1956, 12 satellite tracker stations were deployed marking the beginning of the era of observatory networks  \citep{Whipple1956}. Since then many networks comprising telescopes with a wide range of apertures have been designed and commissioned. Table \ref{tab:Networks} lists selected networks of telescopes that have inspired us during the design process. 

\begin{deluxetable*}{@{}p{4cm}p{2cm}p{6cm}l@{}}
\tablecaption{Selected networks of telesocpes.}
\tablehead{\colhead{Project} & \colhead{Infrastructure} & \colhead{Comments} & \colhead{Reference}}
\startdata			
Probing Lensing Anomalies NETwork (PLANET) & 3 x 1-m & worldwide network that discovered several microlensing phenomena & \cite{Albrow1998} \\
 RoboNET & 3 x 2-m & Hawaii, La Palma and Australia, partially owned by LCOGT & \cite{Tsapras2009}\\
 Robotic Optical Transient Search Experiment (ROTSE-III)  & multiple 0.5-m & globally distributed, aimed at the detection od optical transients & \cite{Akerlof2003} \\
 The Kilodegree Extremely Little Telescope (KELT)  & two sites with wide field 80 mm lenses  & northern and southern Hemisphere sites, dedicated to search for transiting exoplanets around bright stars   & \cite{Pepper2007}\\
 Hungarian-made Automated Telescope Network (HATSouth) & 6 units & unit consists of four 0.18-m f/2.8 optical telescopes on a common mount that have a combined field of view of 8.5\degree x 8.5\degree; dedicated to detect transiting exoplanets & \cite{Bakos2013} \\
 Las Cumbres Observatory Global Telescope Network (LCOGT) & 2 x 2-m, 17 x 1-m, multiple 0.4-m & 7 locations in both hemispheres, dedicated for professional research and citizen science projects & \cite{Brown2013} \\
 Master-II & 7 x twin 0.4-m &  dedicated to observing optical counterparts of gamma-ray bursts (GRBs) &  \cite{Gorbovskoy2013} \\
 RAPid Telescope for Optical Response (RAPTOR) & 2 arrays & optical transients monitoring, spectroscopy & \cite{White2004} \\
 Wide Angle Search for Planets  (SuperWASP)& 2 arrays & array consists of eight wide-field cameras on an equatorial mount, 482 square degrees total field of view, exoplanetary transits & \cite{Pollacco2006} \\
 MOnitoring NEtwork of Telescopes  (MONET) & 2 x 1.5-m & time available to schools, photometry and spectroscopy & \cite{Bischoff2006} \\
 Pi of the Sky & two arrays & multiple telephoto lens and custom CCD cameras &  \cite{Wawrzaszek2010}\\
Skynet Robotic Telescope Network & 9 sites & global consortium of robotic telescopes that use the Skynet job-queuing & \cite{Reichart2008} 
\enddata
\label{tab:Networks}
\end{deluxetable*}

Apart from telescope networks, a large amount of single autonomous telescopes operate all around the globe - both professional and amateur.


\section{Network Description}
\label{sec:NetworkDescription}

\subsection{Remote sites}
\label{ssec:RemoteSites}
\begin{figure*}[htb!]
\begin{center}
\includegraphics[width=\textwidth]{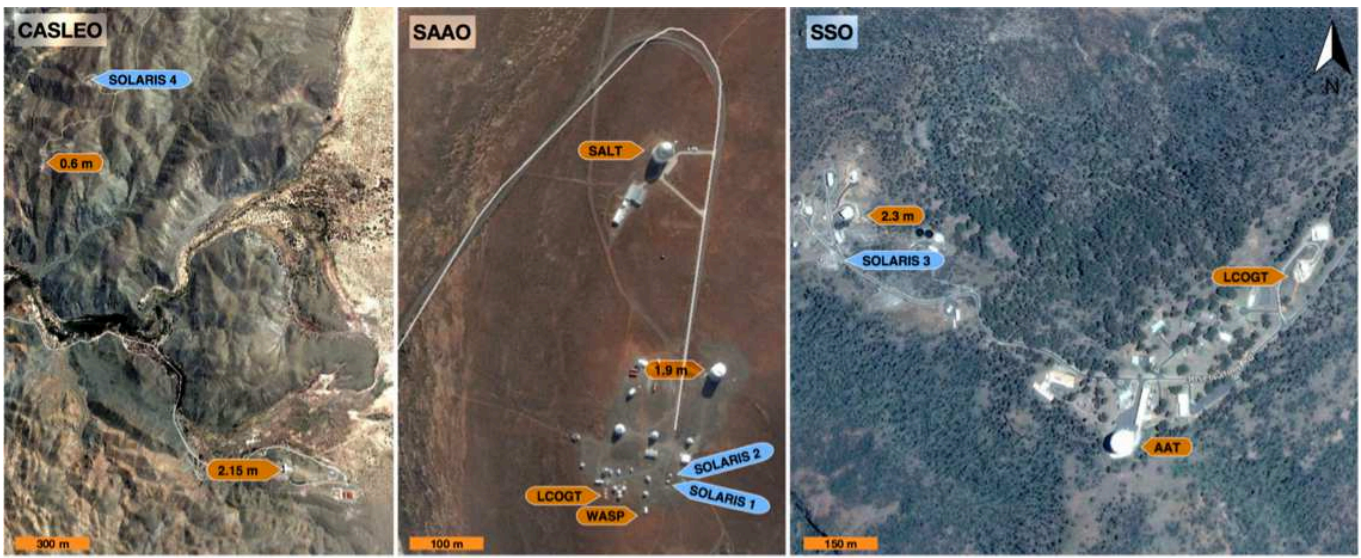}
\caption{Birdseye views of the three observatories with exact Solaris sites marked along with several important waypoints at the South African Astronomical Observatory (SAAO, South Africa), Siding Spring Observatory (SSO, Australia) and Complejo Astron\'omico El Leoncito (CALSEO, Argentina). North is up on all images.  Background image source google.com.}
\label{fig:SolarisMap_setellite}
\end{center}
\end{figure*}

\begin{figure}[htb!]
\begin{center}
\includegraphics[width=\columnwidth]{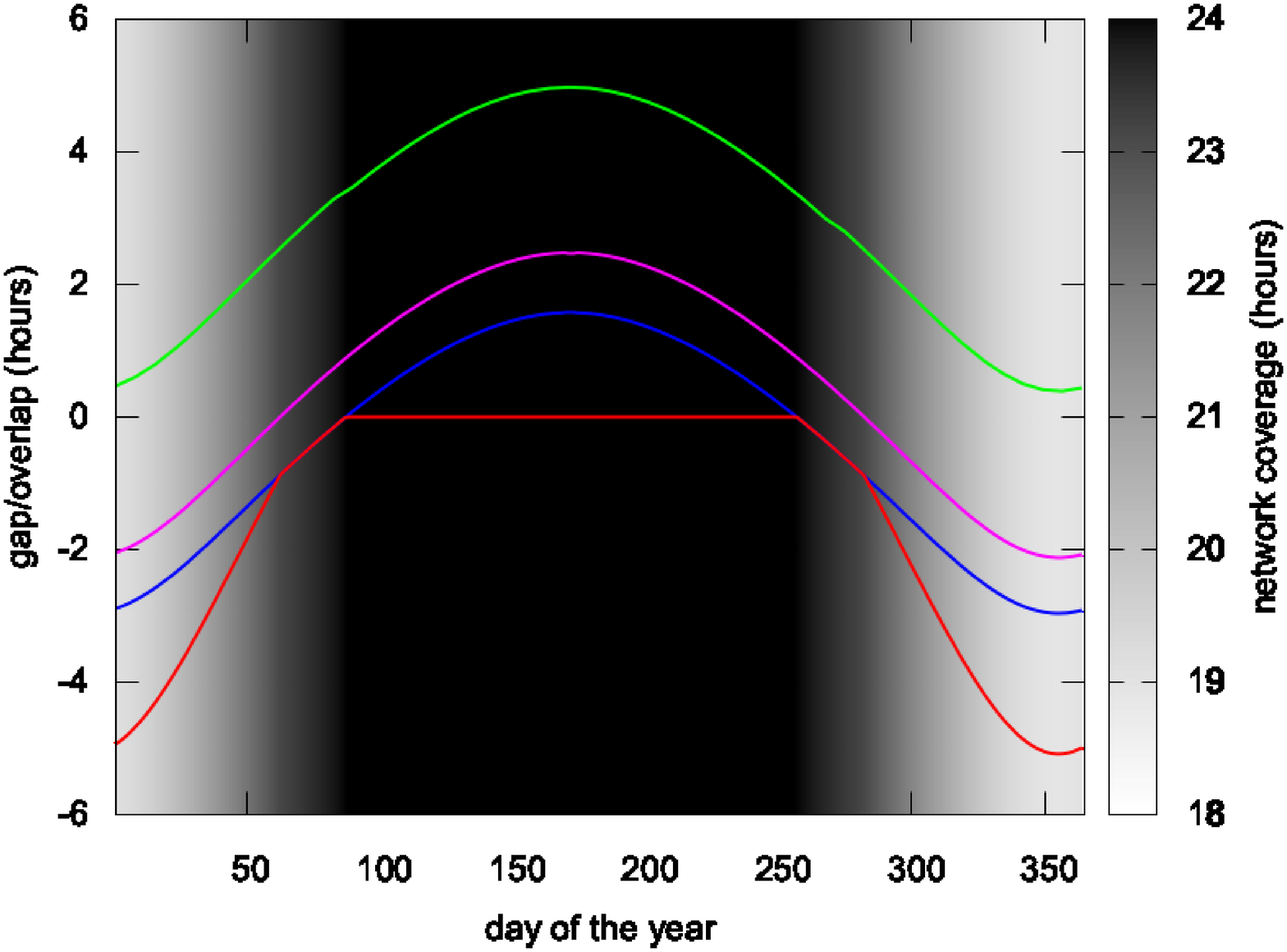}
\caption{Network nighttime coverage throughout the year. Nighttime occurs whenever the Sun is below -18\degree ~altitude.  Plots indicate the number of hours that nighttime between respective sites overlaps (positive values) or has a gap (negative values) per every 24 hours. Magenta, green and blue colors represent the following site pairs: SSO - SAAO, SAAO - CASLEO and CASLEO - SSO, respectively. The red line represents the sum of the gaps in coverages for the entire network. The network covers permanent nighttime from the end of March till mid-September, 46\% of the year. During the southern hemisphere summer the total gap in coverage reaches 5 hours per day. The largest gap occurs between CASLEO and SSO due to the largest longitudinal separation.}
\label{fig:NetworkCoverage}
\end{center}
\end{figure}

The Solaris observatories are located in the Republic of South Africa, Australia and Argentina. Figure \ref{fig:SolarisMap_setellite} gives a graphical overview. All of them lie within less than 1\degree\ difference in latitude. The nighttime coverage is shown in Fig.~\ref{fig:NetworkCoverage}. The plot shows a theoretical result that takes into account only the Sun's position at the respective sites. Actual object observability will be determined by its coordinates and the weather conditions.


The \textbf{Solaris-1 and Solaris-2} telescopes are located on the premises of the South African Astronomical Observatory near Sutherland in the Hantam Karoo at an elevation of 1842 m AMSL. They share the plateau with many other telescopes among which are the Southern African Large Telescope, a station of the Birmingham Solar Oscillations Network (BiSON), the Kilodegree Extremely Little Telescope (KELT-South), LCOGT, Monet, SuperWasp, Master. SAAO's infrastructure is very well developed and managed with very good technical support. 

\textbf{Solaris-3} is located at the border of the Warrumbungle National Park, near Coonabarabran in New South Wales, Australia at 1165 m AMSL elevation. This volcano crater location posed construction difficulties at the same time being a very picturesque area. The observatory is home to the  3.9-m Anglo-Australian Telescope (AAT), Faulkes South, HAT-South, ROTSE, UK Schmidt Telescope (UKST), the The Automated Patrol Telescope. The technical support staff at the mountain is very professional. 

\begin{figure*}[htb!]
\begin{center}
\includegraphics[width=\textwidth]{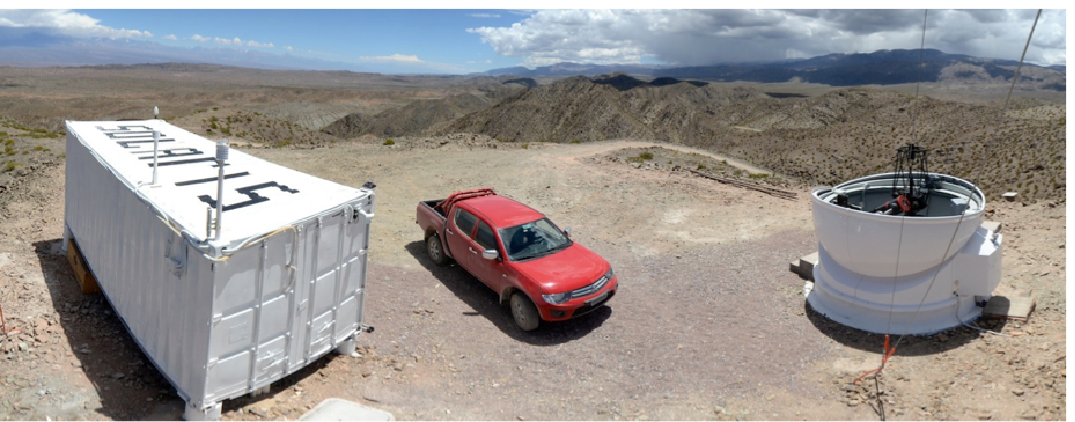}
\caption{Solaris-4 site, CASLEO, Argentina. The site is accessible only with a proper 4x4, especially during and after rainfall. The steel cables visible on the right are guy-wires that support the lightning mast.}
\label{fig:casleo}
\end{center}
\end{figure*}

\textbf{Solaris-4} is located in the El Leoncito National Park. The Complejo Astron\'omico El Leoncito observatory is located in the San Juan province. The site comprises the main observatory buildings including the Jorge Sahade 2.15-m telescope and a remote location at a higher elevation of 2552 m AMSL, 7 km away - this is where Solaris-4 has been built. This site has been particularly challenging in terms of customs regulations and bad road conditions in the area (Fig. \ref{fig:casleo}). The network's headquarters is located in Toru\'n, Poland.

\section{System components}

\label{sec:HardwareComponents}

\begin{figure*}[htb!]
\begin{center}
\includegraphics[width=\textwidth]{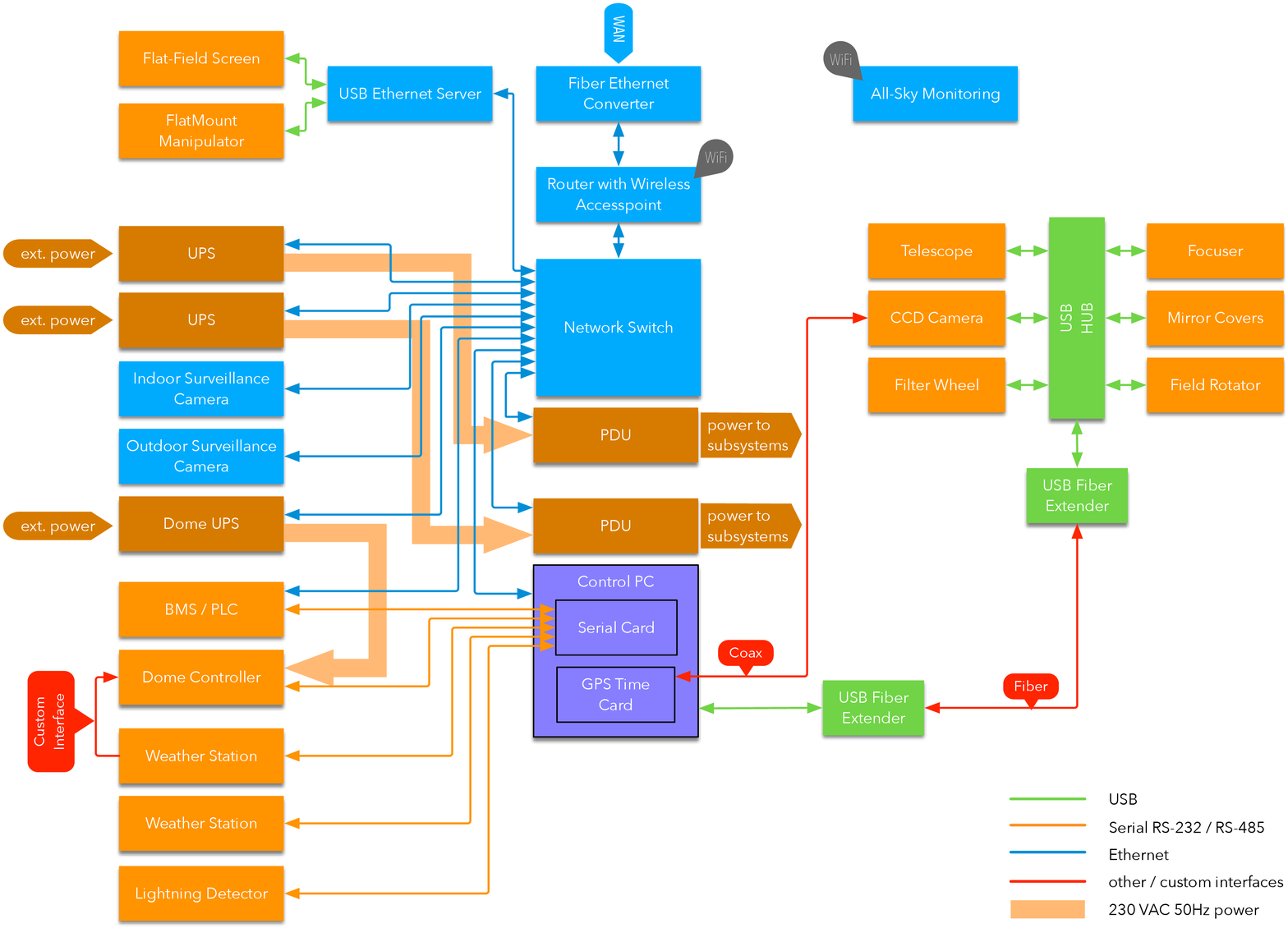}
\caption{Low-level architecture of a Solaris node. The diagram presents an overview of all components used in the observatory along with the most important communication channels that interconnect the elements of the system. A single server is used to control and integrate all subsystems (Tab. \ref{tab:ComputerHardware}). Original graphic from \cite{Kozlowski2014}.}
\label{fig:SolarisOverviewDiagram}
\end{center}
\end{figure*}

\begin{figure*}[htb!]
\centering
\includegraphics[width = \textwidth]{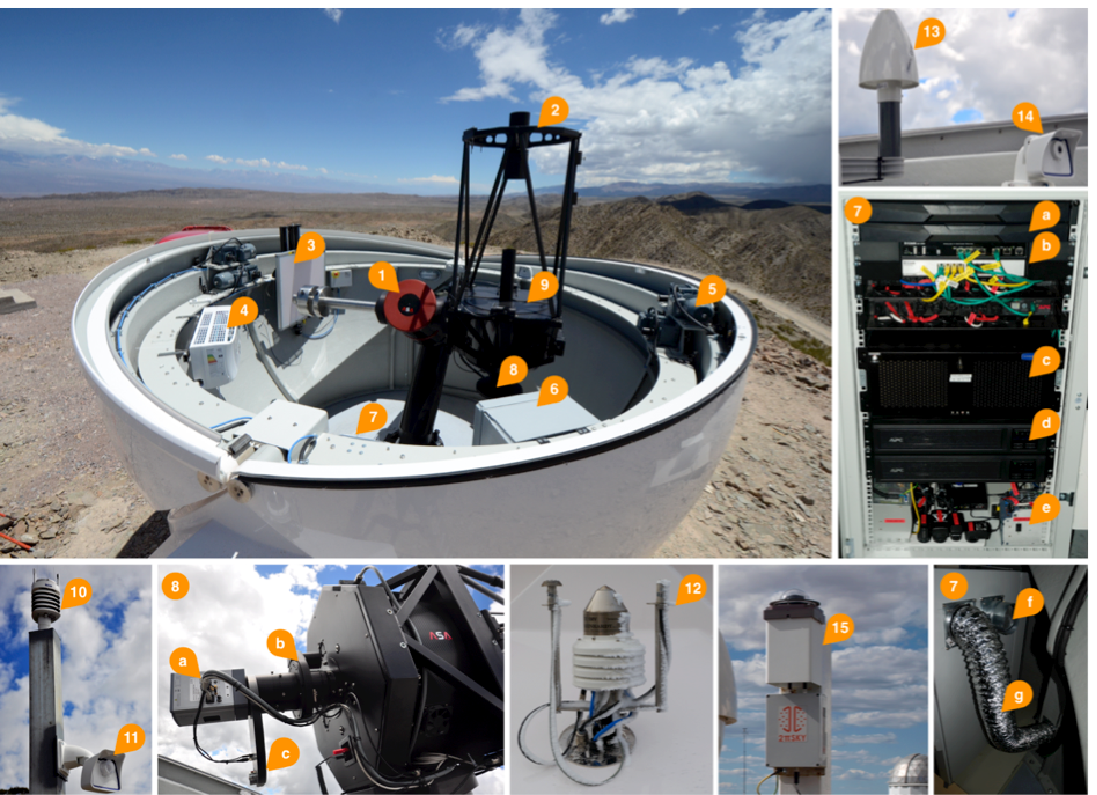}
\caption{Solaris hardware overview. 
1. ASA DDM-160 telescope mount.
2. 0.5-m optical tube assembly.
3. Flatfield screen and FlatMount lifting manipulator.
4. Air-conditioner internal unit.
5. Dome segment motors.
6. PLC HVAC and supervision system.
7. 19-inch rack cabinet:
a - KVM remote console with keyboard and LCD monitor,
b - network router and switch,
c - control PC,
d - UPS units,
e - power supplies and accessories.
8. Imaging train:
a - Andor iKon-L CCD camera,
b - ASA field rotator,
c - FLI filter wheel.
9. Mirror covers.
10. Vaisala WXT-520 weather station.
11. Mobotix external surveillance camera.
12. Reinhardt weather station.
13. Meinberg GPS antenna.
14. Mobotix internal surveillance camera.
15. SBIG all-sky camera with custom \twopiskytm cloud detection module. Graphic based on original from \cite{Kozlowski2014}.}
\label{img:HardwareOverview}
\end{figure*}

The Solaris network's design prerequisites define the detailed requirements that need to be met by the individual observatories. Though the major hardware components are off-the shelf products, they need to be integrated in such a way that the systems operate autonomously. The architecture of the system is presented on the diagram in Fig.~\ref{fig:SolarisOverviewDiagram}. A photographic overview of the components after installation is shown in Fig. \ref{img:HardwareOverview}. The proposed architecture is characterized by the following features.

\begin{itemize}
\item Ability to control the power supply state of all components. In a case, when a hardware reset is necessary, it can be performed by the system automatically or manually by the user. 
\item Power isolation. Sensitive and expensive components, such as the CCD camera and the telescope, are separated from components that are exposed to lightning strikes and power surges, e.g. weather stations, antennas, etc. The use of online UPS units additionally increases the level of power safety. Additionally, all components installed outdoor are equipped with surge arrestors. Fiber is used for data transfer wherever possible.
\item Focus on security. It is crucial for the dome to be closed whenever it is not safe to operate. Our system implements many safety features that supervise the dome. The dome is closed automatically during daytime, bad weather (the rain sensor is hardwired to the controller), in case of a loss of communication with the dome controller (both software errors or hardware problems) and when there is no active internet access on site. All sites but CASLEO offer a stable and reliable internet connection. SAAO even has a backup link. CASLEO, especially during winter, suffers from occasional power losses that eventually lead to major communication problems. Therefore, to be on the safe side, all sites treat the internet access property as one of the operating conditions.
\end{itemize}

Below we present an overview of the hardware setup. A much more detailed  description can be found in \cite{Kozlowski2014}.

\subsection{Astronomical equipment}

\begin{figure}[htb!]
\centering
\includegraphics[width=\columnwidth]{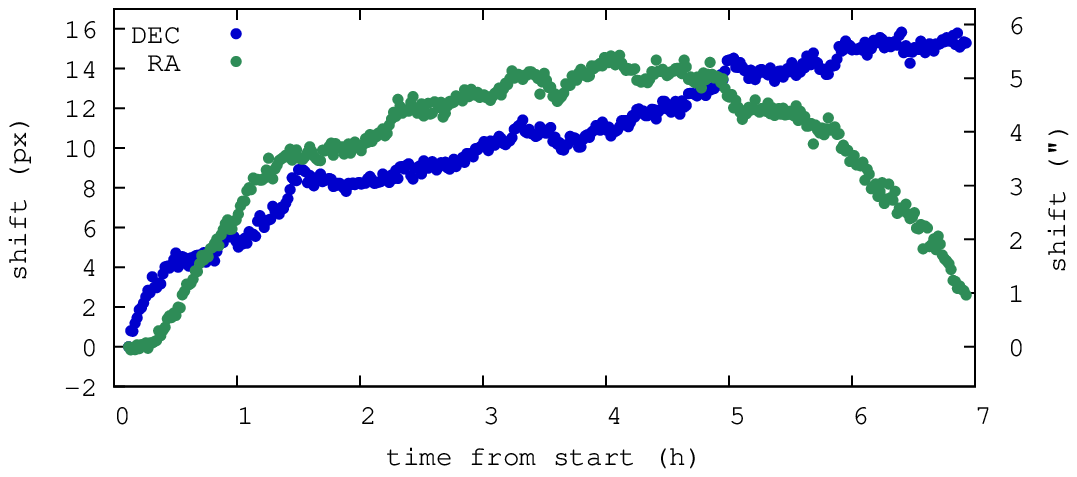}
\caption{RA and DEC position recorded during a 7 hour run on Solaris-2, no position corrections have been made during this time. Wind speed 12-15 km h$^{-1}$.}
\label{fig:TrackingTest}
\end{figure}

\paragraph{Telescope, mount and accessories.}
Astrosysteme Austria (ASA) mounts and optical tubes have been chosen for the project. Solaris-1, -2 and -4 are Ritchey--Chr\`etien telescopes, Solaris-3 is a Schmidt-Cassegrain design with a field corrector, all riding on ASA DDM-160 direct drive mounts. The DDM-160 mounts are installed on modified piers that allow the telescope to observe past the meridian significantly longer than in case of a classical design. The loading capacity is 300 kg. According to the manufacturer's specification the pointing RMS should be better than 8 arc seconds and the tracking precision should be better than 0.25 arc sec. RMS during 5 minutes - all thanks to hight resolution (0.007 arc sec) incremental encoders. Figure \ref{fig:TrackingTest} shows tracking test results. The maximum slewing speed is 13 degrees/s. Unfortunately, the mounts have a USB interface (based on a FTDI chip) making the setup very sensitive to communication errors. The optical tube assembly is fitted with a focuser, mirror covers and a field rotator -- all motorized and controlled via dedicated software -- Autoslew. Autoslew has a graphical user interface and handles the configuration of the mount. The user can control the parameters of the PID controllers (tabulated for different slewing speeds) and even the parameters of filters that are used in the control loop. The pointing model can be created manually or with the help of dedicated software (Sequence by ASA) that automates the process.

\paragraph{Imaging train.}
All four telescopes are equipped with Andor iKon-L CCD cameras that are based on e2V CCD42-40 chips and fitted with four stage thermoelectric cooling that cools the CCD down to -70\degree~Celsius. The camera's shutter is connected to a GPS card that records the opening and closure times of the shutter providing a very precise time stamp that is then saved in the image header. A Fingerlakes Instruments filter wheel with $\phi$50 mm Johnson (UBVRI) and Sloan (u'g'r'i'z') filter is installed as well. All four imaging trains are identical with the exception of Solaris-3 that has an additional field corrector and Solaris-1 that is fitted with a spectrograph.

\paragraph{\'Echelle spectrograph.}

\begin{figure}[htb!]
\begin{center}
\includegraphics[width=\columnwidth]{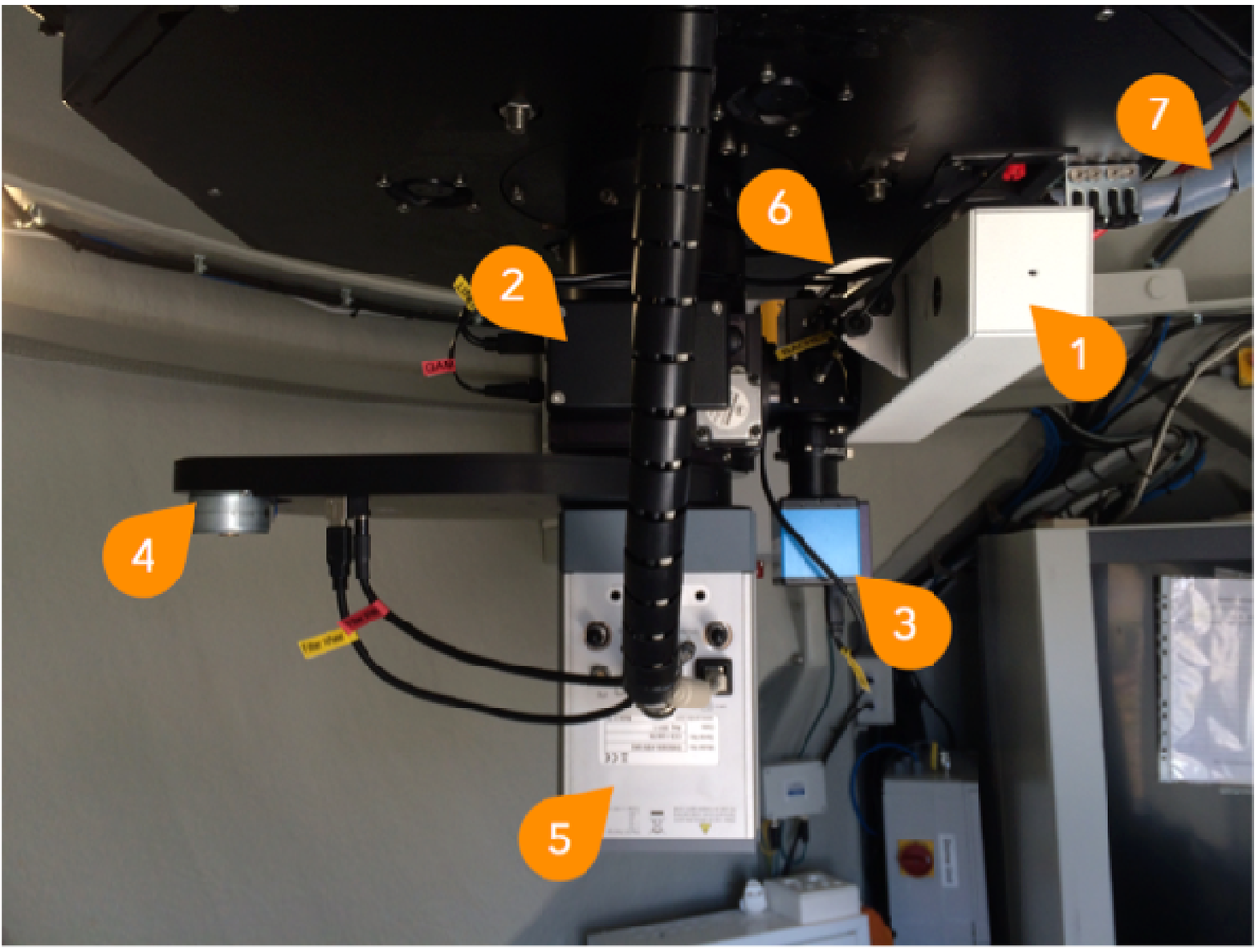}
\caption{Solaris-1 imaging train: 1 -- BACHES spectrograph, 2 -- GAM, 3 -- guide camera, 4 -- filter wheel, 5 -- photometric CCD camera, 6 -- spectroscopic camera (only mounting flange is visible), 7 -- spectrograph control cabling and fiber. Graphic based on original from  \cite{Kozlowski2014}.}
\label{fig:BachesSet}
\end{center}
\end{figure}

In 2013, Baches, a prototype \'echelle spectrograph has been tested on the Solaris-4 telescope \citep{Kozlowski2014}. The spectrograph body is 290x100x52 mm in size and weighs less than 1.5 kg, making it a very compact instrument, even after including the spectroscopic and guide cameras. Both are very conveniently attached to the instrument.  Internally, the instrument consists of a collimator lens, a 63 l/mm 73$\degree$ \'echelle grating, a cross-dispersing diffraction grating and an objective. The instrument is optimized for f/10 input beams and cameras based on the KAF-1603 CCD chips. After a successful test campaign, a final production version of Baches has been installed permanently on the Solaris-1 telescope in South Africa -- Fig. \ref{fig:BachesSet}. To avoid the need of manual instrument changes (photometry remains the main observing mode of the telescope), the imaging train has been fitted with a custom designed guide and acquisition module (GAM). The unit includes an internal flip mirror allowing for remote selection of the desired instrument: CCD camera or spectrograph. The spectrograph setup also includes a remote calibration unit (RCU) that is used to feed light from quartz and thorium-argon calibration lamps to the spectrograph via a fibre. After commissioning, the spectroscopic mode has been thoroughly tested during a dedicated observing campaign \citep{Kozlowski2016}.

\label{sec:spectrograph}

\subsection{Computer hardware}
\label{ssec:ComputerHardware}

Each observatory is controlled by a single server-grade computer fitted with a GPS card and a multi-port serial card (Tab. \ref{tab:ComputerHardware}). Fiber is used where needed  (Fig.~\ref{fig:SolarisOverviewDiagram}) and all components have UPS backup power. Practice shows that USB connections can be unstable, especially when large data throughput is required. To maximize robustness, the following has been taken into account: USB data cables short as possible, respecting the standards (e.g. microUSB and similar have cable length limits), use of tested configurations (e.g. the Icron USB-fibre extender), use of reliable power supplies with proper cabling (cable gauge  adjusted according to the current, voltage requirements and length) and proper grounding of all equipment.

\begin{deluxetable*}{rp{13cm}}
\tablecaption{Computer hardware as of January 2017.}
\tablehead{\colhead{component} & \colhead{description}}
\startdata			
PC  				&  SuperMicro server-grade motherboard, Intel Xeon with 32GB RAM, 200 GB SSD (RAID 1), 12 TB storage \\ 
computer access	& KVM with integrated LCD monitor and keyboard \\
time source		&  high precision PCI Express GPS receiver with a dedicated antenna and external hardware time event capture  \\ 
internet access		&  1Gbit fiber to ethernet converter, 1Gbit router and 24-port 100Mbit ethernet switch  \\ 
power backup 		&  two on-line UPS units: 3000VA and 1500VA  \\ 
power distribution	&  two power distribution units, network enabled \\
power supplies		&  astronomical equipment, incl. telescope with focuser and mirror covers, CCD camera and imaging train components, flat field screen and FlatMount, IP surveillance cameras, weather monitoring devices, emergency lighting
\enddata
\label{tab:ComputerHardware}
\end{deluxetable*}

\subsection{Dome and infrastructure}

\paragraph{Dome.}
Clamshell domes manufactured by Baader Planetarium GmbH have been selected for the project. The 3,5-m domes are made of  fibre-reinforced plastic and are self-sustained structures consisting of a cylindrical dome base and four motorized segments that allow the dome to be opened fully and provide an unobscured view of the entire sky. The sandwiched structure is well insulated. The concrete footing precisely matches the external circumference of the dome and is separated from the telescope's pier base to prevent the transfer of vibrations from the dome onto the telescope. 

\paragraph{Weather stations.}
Real time weather information is provided by two weather stations that measure temperature, relative humidity (redundant), wind speed and direction, precipitation (redundant) and the cloud base. One of the rain sensors is additionally hard-wired to the dome controller for extra security. 

\paragraph{All-sky camera.}

\begin{figure*}[htb!]
\begin{center}
\includegraphics[width=0.49\textwidth]{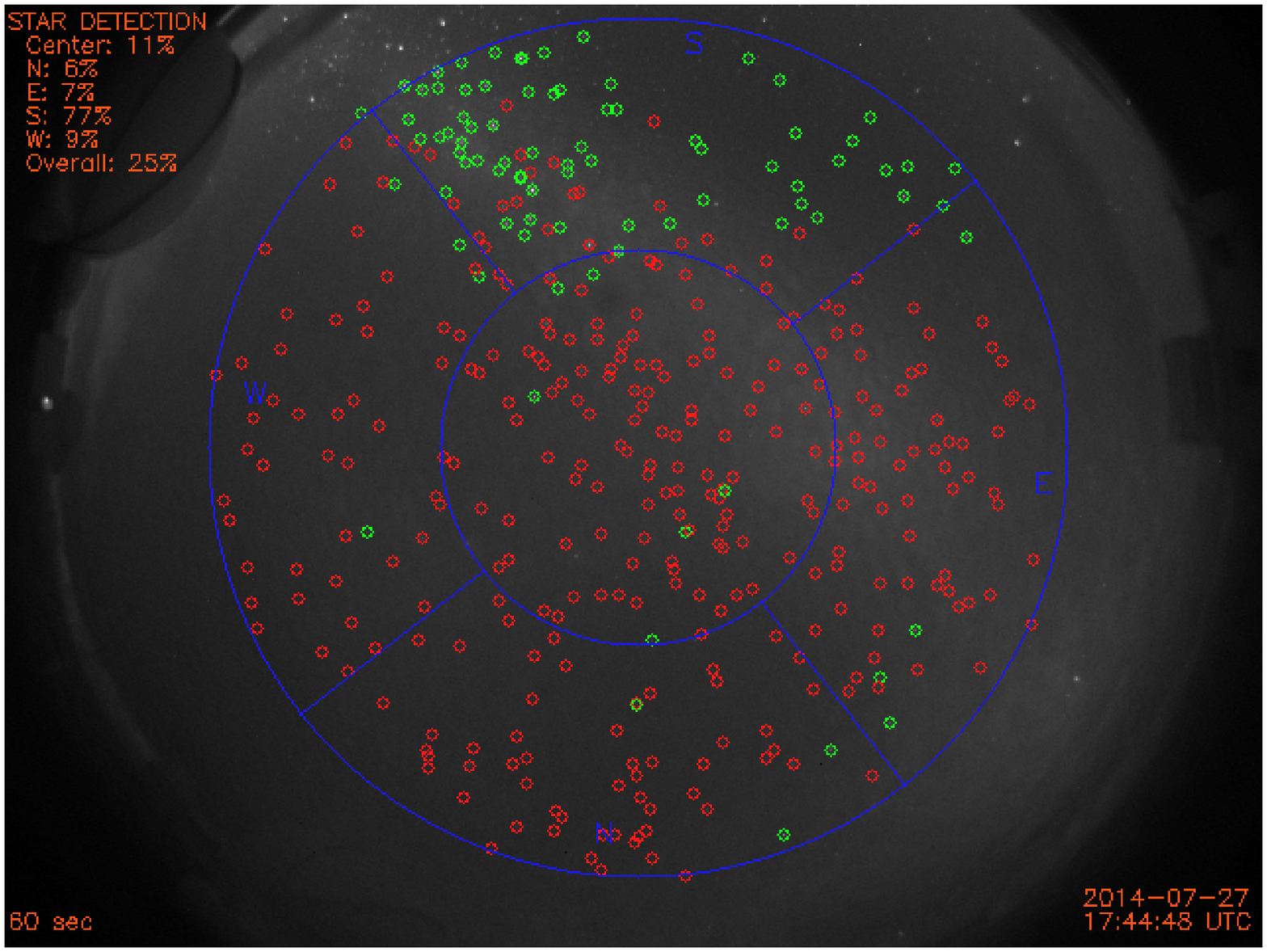}
\includegraphics[width=0.49\textwidth]{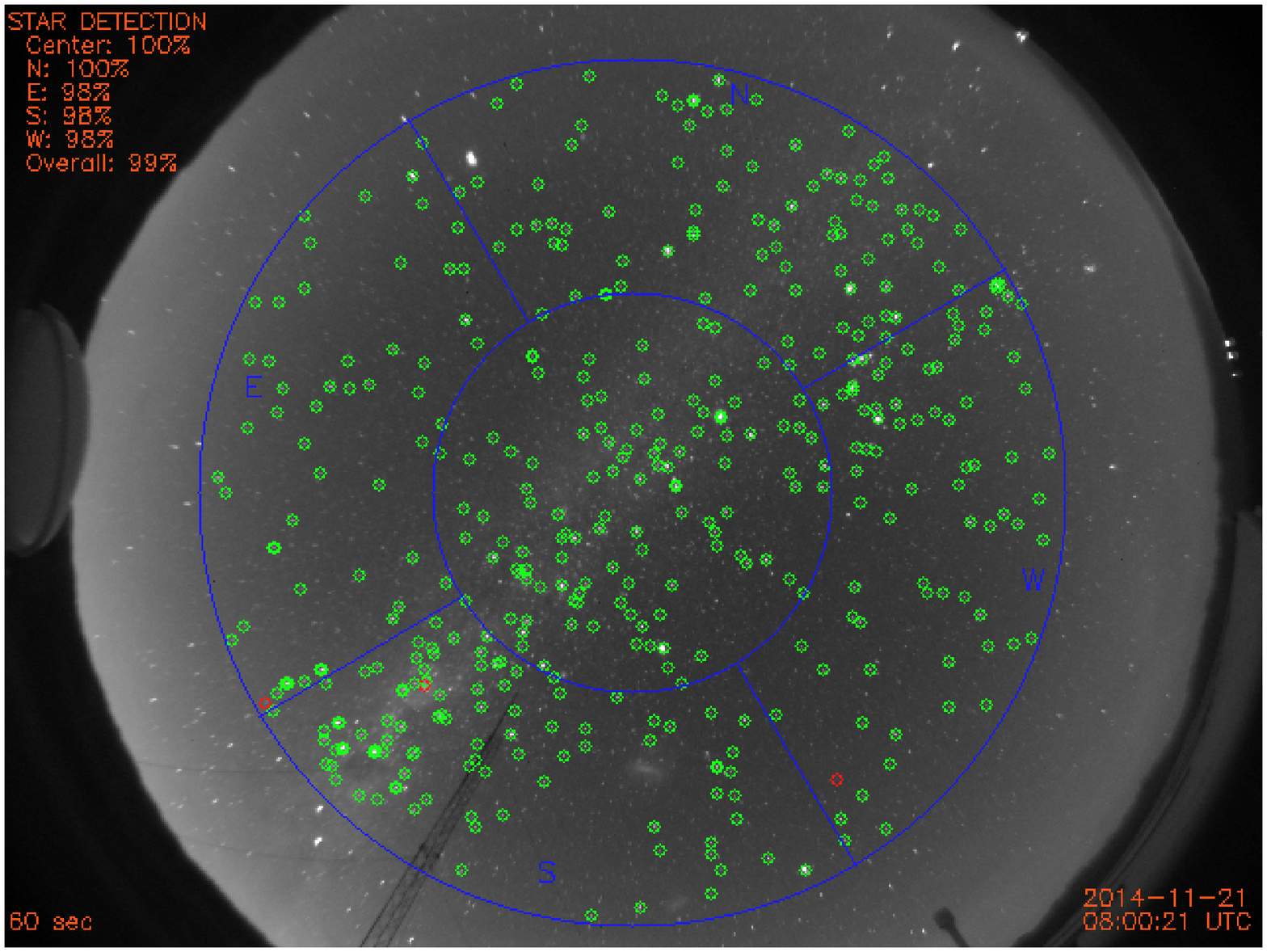}
\caption{Sample all sky images obtained with the all-sky monitoring system installed in Colmplejo Astronomico El Leoncito in Argentina. Each of the 60 second exposures shows different cloud coverages. Green circles denote stars properly detected and identified in the catalog, red circles denote stars obscured by clouds.}
\label{fig:AllSkyImages}
\end{center}
\end{figure*}

To provide precise cloud coverage information all sites have been equipped with an SBIG all-sky camera and a \twopiskytm add-on module that provides the cloud detection functionality. The device analyses the all-sky images and determines the cloud coverage based on photometric measurements. The results are used as input by the control system. \twopiskytm also provides a web interface that gives access to the configuration options and image database (Fig. \ref{fig:AllSkyImages}). The approach is sensitive to high altitude clouds that are not detected by the simple temperature-based  cloud sensors. 

\paragraph{Flatfield screen and FlatMount.}
For camera calibration purposes (nonlinearity, shutter effects and general troubleshooting), all sites have been equipped with 0.6x0.6-m electroluminescent flatfield screens that are mounted on a dedicated manipulator called the FlatMount. The device can raise the screen during calibration and lower it during normal observing so that the sky view is not obscured. 

\paragraph{Surveillance cameras.}

All sites are under video surveillance. IP cameras are installed inside the domes (monochrome, high sensitivity, fisheye lens) and oudoors to overlook the domes (color cameras with narrow-angle lenses). 

\begin{figure}[htb!]
\begin{center}
\includegraphics[width=\columnwidth]{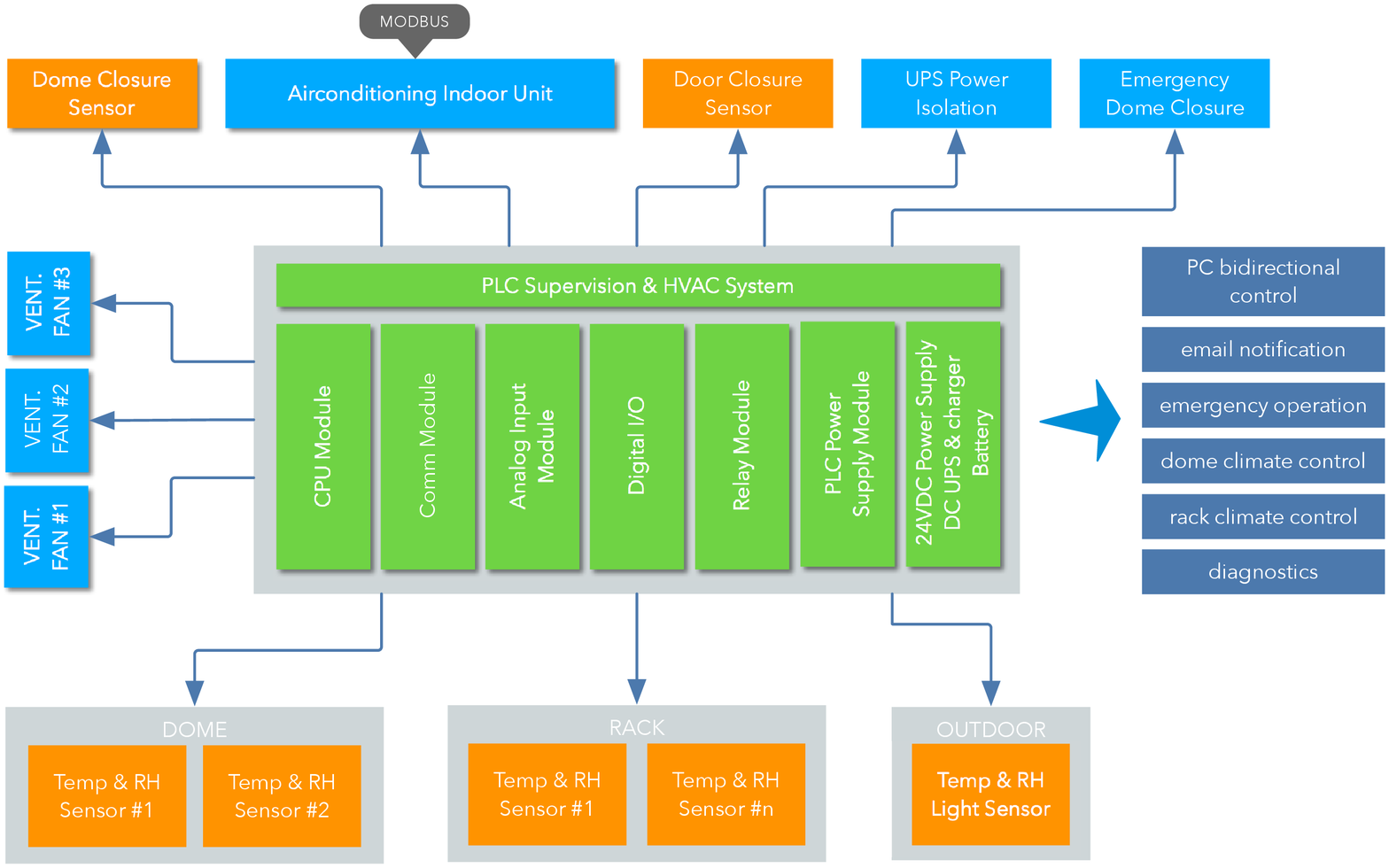}
\caption{Building Management System design overview. \cite{Kozlowski2014}.}
\label{fig:PLCdesign}
\end{center}
\end{figure}

\paragraph{Building management system.}
\mbox{}\newline \owtm~is a PLC-based  building management system that handles the heating, ventilation and air conditioning of the dome, emergency closure and rack cooling. This approach was chosen to popularize industry standards in the world of small astronomical observatories. The design of the PLC system is illustrated in Fig. \ref{fig:PLCdesign}. 

\paragraph{Lightning protection, grounding and electrical system.}
Proper grounding and lighting protection is extremely important in remote, high altitude locations. All Solaris sites have a carefully designed TT-type electrical network with proper grounding and lightning protection based on bentonite, copper rods and deep rock drillings. The Solaris-4 site in Argentina, due to its solitude in the area, has a dedicated 10.5-m lightning protection mast. The mast can influence the quality of photometry, but it's active cross-section is minimal and is therefore not taken into account during observing planning.

\section{Software architecture}
\label{sec:SoftwareArchitecture}
\subsection{Existing solutions}

In the astronomical domain there already exist solutions that target the needs of astronomical observatories, but although they have their usage, an end-to-end solution for managing a global network of robotic telescopes has not been their aim. 

\subsubsection{Communication Protocols and Platforms}
The predominant solution currently used by the device manufacturers is ASCOM \citep{2002SASS...21...39D}, a many-to-many, language independent platform supported by astronomy for Windows-based computers. Smaller in terms of available devices, INDI, Instrument-Neutral Distributed Interface, a library working with POSIX operating systems is also worth mentioning. One of the advantages of INDI was the development of an XML-based protocol. Both, ASCOM and INDI, aim at providing the lower layer of communication with the astronomical observatory components, but focus on the core components, such as: mount, telescope, camera, filter wheel, etc., and not the auxiliary components like: power distribution units, uninterruptible power supplies, weather stations, environmental monitors, etc. that are required for robotic operation of an observatory. This means that currently most of the hardware that is required for an observatory to work in a fully robotic mode has custom communication mechanisms and the main task in programming such mode lays on integrating these hardware components into a system.

\subsubsection{Management Systems}

On top of the aforementioned platforms higher level solutions are based. Describing them is not the purpose of this paper, but it may be valid to point their characteristics and compare them to the requirements of Project Solaris. Software such as MaximDL\footnote{http://diffractionlimited.com/product/maxim-dl/} \citep{2004S&T...108d..91W} or ACP\footnote{http://acp.dc3.com/index2.html} \citep{2000IAPPP..80...13D} works as one-stop program that is installed on a machine connected to all the devices the observatory controls and relies on and provides the means of interacting with them. In Project Solaris the required level of operation has been defined as such that allows the software to work as-a-service/as-a-daemon and retrieve the observing plans, execute them if possible and transfer the resulting data to the headquarters located in Toru\'n. This means that the software on-site should work as a client to the service that feeds it with observing programs, but a client that is fully aware of the environmental situation and can decide on executing an observation under safe local conditions. It must not be human-operated with the exception of hardware testing or functionality/performance analysis.
Remote Telescope System, RTS2, a Linux-based software is a good example of software that can allows for operating a remote observatory, automate the process of executing observation programs, etc. \citep{2010AdAst2010E..88K}. However, at the time of software selection for Project Solaris RTS2 did not have most of the hardware components described in the previous sections supported. Adding to that, ASA mount with its accessories (i.e. focuser, mirror covers and field derotator), came with a Windows-based software system, Autoslewª and was fully available through ASCOM.

\subsection{Our approach}

In Project Solaris after an in-depth research and analysis of the available options a dedicated system has been developed with the aim to satisfy the requirements described in section \label{sec:SystemComponents}. The platform of choice was ASCOM as most of the core components had fully-functional drivers on this platform and all were accessible on Windows.

\begin{figure}[ht]
\centering
\includegraphics[width = \columnwidth]{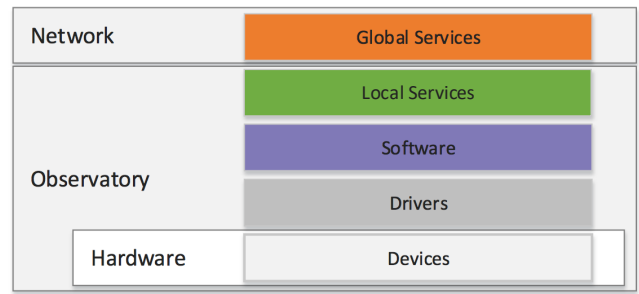}
\caption{The architecture proposed for Project Solaris that defines the boundaries and specifies the control domains.}
\label{fig:architecture}
\end{figure}

Relying on that a layered architecture was proposed as in Fig. \ref{fig:architecture}. There have been identified 5 layers required for robust operation of a single observatory and connecting it to the global network. They will be described in detail further in the paper. Starting with ASCOM (as the necessary component to communicate with components from ASA) we identified software requirements for developing a solution. Microsoft Windows Server platform was selected as an operating system of choice. Thanks to Microsoft Imagine (formerly DreamSpark)  subscription for STEM institutions all the components such as operating systems, databases, toolkits and platforms were freely available. The overall software components are presented in Table \ref{tab:soft_components}.
	
\begin{deluxetable}{rl}	
\tabletypesize{\scriptsize}
\tablecaption{Software components for creating architecture for Project Solaris.}			
\tablehead{\colhead{System Component} & \colhead{Selected Option}}			
\startdata			
Operating System	&	Microsoft Windows Server 2008 R2	\\
Programming Platform	&	.NET, Microsoft Robotics, Microsoft Azure	\\
Programming Languages	&	C\#, C++, ES6, TypeScript	\\
Standards \& Protocols	&	FITS, XML, Json, OData, Web Sockets	\\
Security	&	X.509, OAuth 2.0, OpenID	\\
Persistence Layer	&	Microsoft SQL Server 2008 R2	
\enddata			
\label{tab:soft_components}			
\end{deluxetable}			

Devices were to be accessible through a low-level interface via asynchronous drivers. These were consumed by software that contained the logic to operate securely the component, allowed for plug-in based utilization of different drivers for devices of the same kind, and performed all the required operation for fault-handling: when allowed Ð fault tolerance, when disallowed graceful failure. This layered architecture is depicted in Fig. \ref{fig:architecture}. To cover for these features, Microsoft Robotics platform was selected. At the time it was the only mature platform to provide for creating robots. And in the case of Project Solaris the requirement for an observatory to operate robotically was mandatory.

The main components of Microsoft Robotics that provided for asynchronous operation and decoupling of the devices operation were CCR and DDS (Concurrency and Coordination Runtime and Decentralized Software Services, respectively). They formed a lightweight, easy-to-program environment for creating services (DDS) that were performing their operation without blocking (CCR) which means that all the components could be orchestrated together rather than operated separately until they finish the operation. Furthermore it meant that the services run in isolation (DDS) and all the problems could be diagnosed and addressed separately and the system was able to save time of operation even if it was only seconds per few operations, as eventually it adds up to hours.

\begin{figure}[htb!]
\centering
\includegraphics[width = \columnwidth]{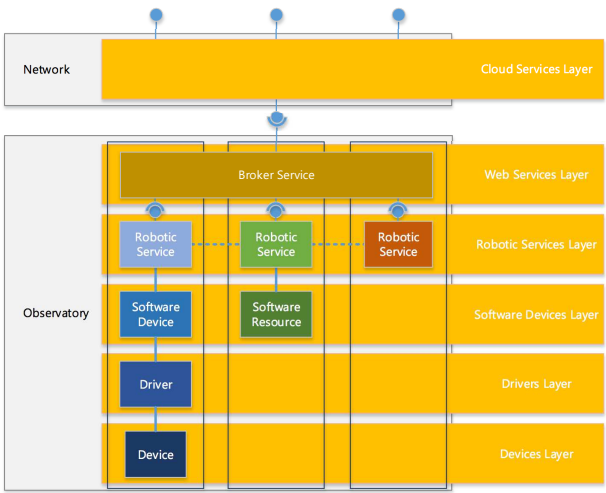}
\caption{ Final architecture of a system for Project Solaris.}
\label{fig:final_architecture}
\end{figure}

Finally, the architecture was updated to include this robotic service as a separate layer as in Fig. \ref{fig:final_architecture}. The figure shows three possible kinds of robotic services. They represent the three identified elements:
\begin{enumerate}
\item A service that relies on a hardware component Ð the most common element, e.g. mount, camera, etc.
\item A service that relies on a software component Ð for example a service that communicates with the database, or a service that performs auxiliary decoration of the metadata for the captured frame.
\item A service that only interacts with other services Ð this will be explained in more detail further in the text, but basically it describes a hub, a service that gathers other services of a specific kind and provides an emergent functionality on top of them.
\end{enumerate}

One can think of the ladder of layers as: a device is accessible from the system through the driver. The driver is consumed by a software device that does not know the specifics of the device, but knows its functionality and can operate it securely and safely. A robotic service utilizes software device to provide the means of isolated, asynchronous and fault-tolerant operation as well as communication with other robotic services. A robotic service is also exposed securely to the higher level, external cloud service by a broker service. The cloud service itself represents the unifying component (a server) for all observatories (clients) in a global client-server architecture. 
Moreover, cloud service also has the means for interacting with even higher level components. It is designed for providing plugins of operation (i.e. substituting schedulers, providing specific persistence and logging mechanisms) as well as allowing for up-to-date data presentation and even user interaction. This level of communication, rather than communicating directly with the observatories has a benefit in that the direct links to the observatories are a high-value resource and they should not be extensively used. 

\subsubsection{Single Observatory}
From Fig. \ref{fig:final_architecture} one can see that a single observatory contains multiple layers as well as is modular by design. Moreover, the requirement to perform its operation robotically, a single observatory consists of multiple various components: hardware-bound, software-resource-bound and operation-only components implemented as isolated services. Apart from the so-called core components that provide the functionality to actually perform the observational task the wealth of components is required mainly to secure the operation of a single observatory and protect the equipment. The full list is presented in Table \ref{tab:protocols}. Even though the connection to the observatory may be established using Ethernet, the actual network is a private one and is bound within a single observatory where possible. For example, the access to the UPSes and PDUs is accomplished with a network connection, but the connection doesn't go outside the subnet. It means that the actual software robotizing the work of a single telescope is deployed at a site. What is, however, deployed remotely from the site, is the global scheduler which allows for serving as an overlord tasking all the telescopes.

\subsubsection{Software Components}
The components in each of the observatories in Project Solaris were divided into core components, basic devices that are required to provide the main functionality of the observatory: generating images; sensor components that are required for the observatory to receive timely information about the local environmental conditions and to protect the system; managing components that are components combining other components to provide a functionality on top of them.

\begin{deluxetable*}{lcccc}		
\tabletypesize{\scriptsize}				
\tablecaption{Solaris Network site summary}									
\tablehead{\colhead{Component} & \colhead{Manufacturer} & \colhead{Interface}& \colhead{Protocol} & \colhead{Driver}}									
\startdata									
Mount	&	ASA	&	USB	&	n/a	&	ASCOM	\\
Focuser	&	ASA	&	USB	&	n/a	&	ASCOM	\\
Mirror Covers	&	ASA	&	USB	&	n/a	&	ASCOM	\\
Field Derotator	&	ASA	&	USB	&	n/a	&	ASCOM	\\
Dome	&	Baader Planetarium	&	Serial Port	&	text-based	&	custom	\\
PDU	&	Neol, APC	&	HTTP /Telnet	&	n/a	&	custom	\\
Camera	&	Andor	&	USB	&	n/a	&	custom	\\
Filter Wheel	&	Finger Lakes Instruments	&	USB	&	n/a	&	custom	\\
FlatField Screen	&	Alnitak	&	Serial Port	&	n/a	&	custom	\\
FlatField Mount	&	Unisar	&	Serial Port	&	n/a	&	custom	\\
GPS	&	Meinberg	&	PCIE	&	n/a	&	custom	\\
UPS	&	Eaton, APC, IntelliPower	&	TCP/IP	&	SNMP	&	custom	\\
Weather Station (dome-hardwired)	&	Reinhardt	&	Serial Port	&	text-based	&	custom	\\
Weather Station	&	Vaisala	&	Serial Port	&	text-based	&	custom	\\
ObservatoryWatch	&	Cilium Engineering	&	Serial Port	&	text-based	&	custom	\\
2PiSky	&	Cilium Engineering	&	TCP/IP	&	text-based	&	custom
\enddata									
\label{tab:protocols}									
\end{deluxetable*}									

\subsubsection{Core Components}
\paragraph{Mount and accessories.}
ASA DDM160 mount is accessible via Autoslew, a proprietary window application that runs on Windows. The application also starts an ASCOM server allowing for a programmatic access to the mount. Alongside, an additional window application, ACC, is started that allows for manipulation of focuser position, telescope mirror covers and, if present, field rotator. The devices themselves are connected to the computer via a USB cable. 
The application is easy to use, but in Solaris we want to have the possibility to use the devices as windowless services that was impossible for the mount. Autoslewª had to be always open. It required administrative privileges to work which required to set up a dedicated, administrative account that would have its window session always open. 
\paragraph{Dome.}
AllSky Baader Dome is accessible from the system via a serial port connection and a simple one-line description protocol to inform about the state of the four segments: if they are open, closed or in the intermediate state. The domes have also implemented a safety mechanism when used in so-called automatic mode that requires a computer to request the domeÕs state every period of time. If the request is not issued Ð the dome closes. 
\paragraph{PDU.}
Programmable Power Distribution Units, or PDUs for short, apart from being power sockets for all the other devices provide a way to include the information of the power distribution in the system. Programmable PDUs also allow for setting up each PDU outlet state. In Solaris-1, Solaris-2, and Solaris-4 Neol PDUs are used that have their built-in http server which allows for viewing and modifying outlet states. Solaris-3 utilizes an ACP PDU that is accessible via Telnet. As there were no representation of such a device in ASCOM for Project Solaris there was a custom driver implemented.
\paragraph{Camera.}
Project Solaris uses in all observatories Andor iKon cameras connected through a USB cable. Andor provides multiple driver implementations in C, C++, and in .NET (C\# and Visual Basic), yet all the implementations are simple wrappers upon C language library. The C driver gives a list of possible methods to be invoked by the camera. We have developed our own implementation of a higher level, object-oriented C\# driver wrapping the C one that allows for the discovery of the camera connected to the computer, initializes all the required features and provides means of finalization and disposal.
\paragraph{Filter Wheel.}
Finger Lakes Instruments (FLI) CFW-3-12 filter wheel is connected with a USB and accessed through a C library upon which a C\# wrapper was created. 
\paragraph{Flatfield screen and its mount.}
The observatories for Project Solaris were designed to be compact units. It, however, prevented from placing a flatfield screen high for the telescope to look directly at the screen. Because of that a movable flatfield screen mount was designed for the setup that could move up when the dome is open, but normally would be in its bottom position for the dome to safely close. Both of the devices are connected with a USB cable and programmed using dedicated drivers. In Project Solaris every night sky flats are taken and the flatfield screen has its purpose in determining the camera shutter model and camera linearity model.
\paragraph{GPS.}
In Project Solaris Meinberg GPS is used to receive signals from the camera that inform about the precise times of shutter opening and closing. Meinberg provides a C library to intercept the signals from the device upon which a C\# implementation has been developed for Project Solaris. The GPS card records times of two events: end of shutter opening and beginning of shutter closing. Therefore, the precise mid-expoure time can be computed. Both events are saved in the FITS' headers.

\subsubsection{Sensor Components.}
Sensor components are another set of components in Project Solaris. As the core components focus on the functionality of the system, sensors, on the other hand, focus on securing the systemÕs operation by constantly monitoring the environmental conditions. Redundancy of the sensors is one of the sought features that would help in keeping the system working even if one of the sensors would break. 
The information from the sensors is stored in a local database with insert times from 15 seconds to 5 minutes. That allows to identify the cause of the problem when it occurs even after it happened.
To incorporate the notion of good or bad conditions sensor components are limited in the way that every property that they describe can have limits. When the values of the sensors are in the range of the limits a component reports good observing conditions. When, however, the values exceed the limits Ð the component will report bad observing conditions. Limits for any component can be specified in configuration files and using .NET built-in reflection mechanism it is robust enough to work independent of the value type.
\paragraph{UPS.}
Another, after PDUs, set of devices determining safety in terms of power are uninterruptible power systems, UPSes. UPSes used in Project Solaris are accessible via Ethernet providing an IP address and satisfying SNMP protocol. 
\paragraph{Vaisala Weather Station.}
Vaisala is the primary source of information about the ambient conditions. It is connected via a serial port and sends messages in a simple text format. 
\paragraph{Reinhardt Weather Station.}
Reinhardt Weather Station is the second source of information about the ambient conditions. Similarly to Vaisala, it is connected to the main computer with a serial port and provides the data in a easy to parse text format.
\paragraph{ObservatoryWatch.}
ObservatoryWatch is a system that is a heterogeneous hub for various sensors installed inside and outside the dome built upon an industrial grade PLC that provides low-level security mechanisms for the observatory. ObservatoryWatch is connected to the computer via a serial port and provides a simple text-based protocol accessed through a dedicated C\# library.
\paragraph{2$\pi$Sky.}
All sky camera used in Project Solaris, 2$\pi$Sky, is connected to the system via an IP address over the Wi-Fi. It provides two types of messages: an image for the visual presentation of data and a text-based information set about the quality of a priori defined segments of the sky. 
\subsubsection{Managing Components}
These components, called Hubs, do not represent devices, but combine other services to provide an emergent functionality of the services used. They rely on the security mechanisms already implemented in the core components to manipulate them and their state and utilize the information from sensor components to receive the information about the environmental conditions at the observatory.
\paragraph{Device Hub.}
This hub connects to and starts if not yet started all the core components. Above that, Device Hub is responsible for safety and order, if necessary, in which the components are started. Device Hub also reflects the state of each of the components within its own sate.
\paragraph{Sensor Hub.}
Sensor Hub governs the lifecycle of software componentsÕ services. Beyond that Sensor Hub has two tasks: to communicate with a local database to store the information from the sensors with a specified, usually between 15 seconds and 1 minute, period and to include in its own state the synthetic information about the possibility to observe. 
\paragraph{Observatory Manager.}
Observatory Manager is a component performing the core work in the observatory. It connects to all the other hubs and dispatches the weather information from sensor hub to all the others, receives the observing plans and delegates the tasks for all the other hubs.
\paragraph{Observing Hub.}
This hub is responsible for performing the observations. It receives the information about the current observing conditions from Observatory Manager and incorporates the logic to utilize all the core components in the specified order to obtain the requested image.
It is important to say that bias and dark images are no different from a regular astronomical observations in the technical sense. The Camera ServiceÕs (and the driverÕs) methods for executing exposure includes the information whether to open the shutter performing so-called light frames or to keep it closed performing dark frames.
\paragraph{Focusing Hub.}
Focusing Hub performs automatic focusing for the system. It takes into account the focuses that are stored in a local database, distance to the latest best focus for a specified filter. The procedure also includes the temperature at the observatory and if there is a change in the temperature above \textbf{2.5} degrees Celsius Ð a new best focus will be issued. 
\paragraph{Flatfielding Hub.}
Project Solaris uses sky flat fields for the data reduction. These observations are done during the so-called civil twilight, that is prior to the start of the observing night in the evening and after the observing night in the morning. Flatfielding differs from typical optical observatory observing that it must neglect the light sensor readings of ObservatoryWatch. To protect the camera NOVAS library is used to calculate the exact position of the Sun and during the flatfielding procedure and the telescope is pointed to the opposite position.

\subsubsection{Observatory Operation}
A single observatory in Project Solaris works 24/7. That means it needs to have the knowledge of its position, current time, be able to compute the observing night start and end times including twilight times and duration for calibration, and finally have information about the environmental conditions. GPS gives the information about the precise geographic location of the observatory as well as precise measure of time. In Project Solaris United States Naval ObservatoryÕs NOVAS library \citep{Kaplan2012} is used to calculate the general observing night and twilight times and durations. NOVAS also provides the means of observation plan items final check prior to starting the observation.
\paragraph{Observing Plan Execution.}
The overall procedure for observing plan execution is presented in Fig. \ref{fig:Workflow}. Prior to the twilight before the observing night calibration images, i.e. bias and dark frames, are taken. During twilight sky-flats are performed for all the available filters. After the end of the observing night, during morning twilight again the sky-flats are performed and finally darks and biases are taken.
The observation loop is started at astronomical dusk. The Observatory Manager requests the observation plan from the global servicesÕ queue. After that the scheduled observations are analyzed in order to identify the filters that will be used and for them focuser positions are checked. If there is no focuser position for a specified filter as well as temperature focusing observation with high priority is added to the schedule. This loop is repeated until dawn.
\paragraph{Local Data Persistence.}
There are two types of data persisted at the observatory: observational data that is images in FITS format after every exposure and environmental data persistence. Observational data is stored on hard drives in the file hierarchy. Environmental data is stored in a SQL database to be queried when an exceptional situation happens to investigate the reason of its occurrence.
\paragraph{Observations Persistence.}
Every night observations are stored locally on a dedicated high-volume hard-drive in a simple structure: a folder with the name that is year, month day of current UTC date in a format YYYY-MM-DD. In the folders images from each observing nights are stored as they are generated. As all the metadata is stored in the FITS header the files are self-explanatory in terms of location of the image acquisition, position of the object/field of interest, time of acquisition and conditions of the acquisition. After every observing run the images are transferred to the headquarters vault in Toru\'n.
\paragraph{Environmental Information Persistence.}
The information about the sensor readings from all the sensor components in Solaris are stored locally in a Microsoft Sql Server database, 2008 R2 server. This allows for querying the data to obtain the information about the environmental conditionsÕ change in longer time spans to investigate the reasons of observatory system observation gaps.
\paragraph{Logs and Notifications.}
Each of the software components also logs its operation. The so-called logging sinks can be configured for each component separately, but typically the logs are written in rollback-buffer flat files next to the process running. Their role is to provide administrative information and main purpose is to investigate the causes of crashes Ð if they happen. 
Moreover, the services are also monitored by yet another process that continuously performs consistency checks. The process is also responsible for sending e-mail notifications to the administrator group if there is a system failure that cannot be resolved by the system itself.

\subsubsection{Network of Observatories}
The observatories in Project Solaris work in a network which allows for quick full light-curve coverage. In terms of architecture it follows the prescription shown in the Fig. \ref{fig:final_architecture}. Every observatory is independent in terms of how it operates and if the environmental conditions allow for operation. It is however the cloud service layer that defines what and when will be observed. In this view an observatory is a client to the cloud service that it registers to and requests instructions on what to observe in terms of observing plans.
\paragraph{Observing Plans Preparation and Distribution.}
In Project Solaris there is a predefined set of approximately 300 objects to be observed, although the set is open. Nevertheless, as the targets are known to be mostly detached eclipsing binary systems the scheduler is optimized to prepare observing plans focusing on the eclipses. 
Upon the request from an observatory the scheduler takes into account a set of so-called observing plan rules that comprise both scientific and technical long-term and established requirements. For instance apart from the eclipse time there are the typical safety requirements: position of the Sun and Moon, and horizontal altitude. Moreover, each object can be assigned a priority that can elevate its observations in the queue. It must be noted, however, that the environmental conditions are further reevaluated at the observatories including short-term weather and environmental conditions.
The request for a new plan originates at the observatory and technically is performed utilizing dedicated Microsoft Azure Queues. This allows for secure and efficient data transfer around the world also relying on the SLA\footnote{Service Level Agreement} levels from the operator. 
\paragraph{Data Transfer And Persistence.}
There is a difference between transferring observing programs and transferring the results of observations. Of course, the structure of the data is different, yet here we would like to emphasize the importance of these. If an observing program is lost or obsolete the observatory can request for another one at almost no cost. With the generated data it is different. These must not be lost, must not be tampered with and as soon as possible should be backed up in some way. 
It is worth noting that when a technology for implementing this feature to the Solaris system was being selected BitTorrent Sync was in its infancy and did not meet the requirements. Now it has evolved in a mature library (renamed to Resilio), and although we have not tested it in any way it may be a great tool for secure and resilient data transfer. Nevertheless, at the time for Project Solaris we have decided to develop our own tool for transferring the data.
There are two dimensions in which the data transfer operates: 1) transfer node types and 2) data type. Node types represent the storage elements of the transfer network. There can be:
\begin{enumerate}
\item Source nodes -- the ones that contain the data to be sent, but do not receive any. In the system these are the observatory computers, where the generated data is being stored.
\item Transfer nodes -- the ones that receive the data and are sources of data for further nodes. 
\item Backup nodes -- the ones that receive the data and may be sources of data for further nodes. 

\end{enumerate}The difference between transfer and backup nodes is in persistence: backup nodes never delete any data that is sent to them, whereas transfer nodes delete the data only when the data up in the chain (i.e. farther from source) is deleted or if it is fully stored on the backup node. Source nodes shall delete the data only if it is completely and securely stored on the backup node. Completely here is related to a single data file. For that reason there is a certainty that a data file will never be deleted from any of the nodes in the chain until it is completely persisted on the backup node.
In the developed system a backup node may also serve as a source node for another backup node. Because of that there is a full mirroring of the data persisted in the backup.
The second dimension, relating to the data types gives the distinction between:
\begin{enumerate}
\item Data Ð that is the actual FITS files generated during observations. This type is stored in a regular file structure.
\item Metadata Ð that is the information about the path to the file, name of the file, size of the file and create/update times. This type is stored in a database in a dedicated table.
\end{enumerate}
Data is required for obvious reasons. Metadata serves as the memory information when deciding on transferring, updating and deleting a file from a node.
The nodes provide and consume a resilient, multiplex, two-way \mbox{WS-*} services secured by commercial certificates. To secure the transfer packets from errors a CRC sum is provided with each packet.

\section{Operation}
\label{sec:Operation}

\subsection{Weather statistics}

\begin{figure*}[htb!]
\includegraphics[width=\textwidth]{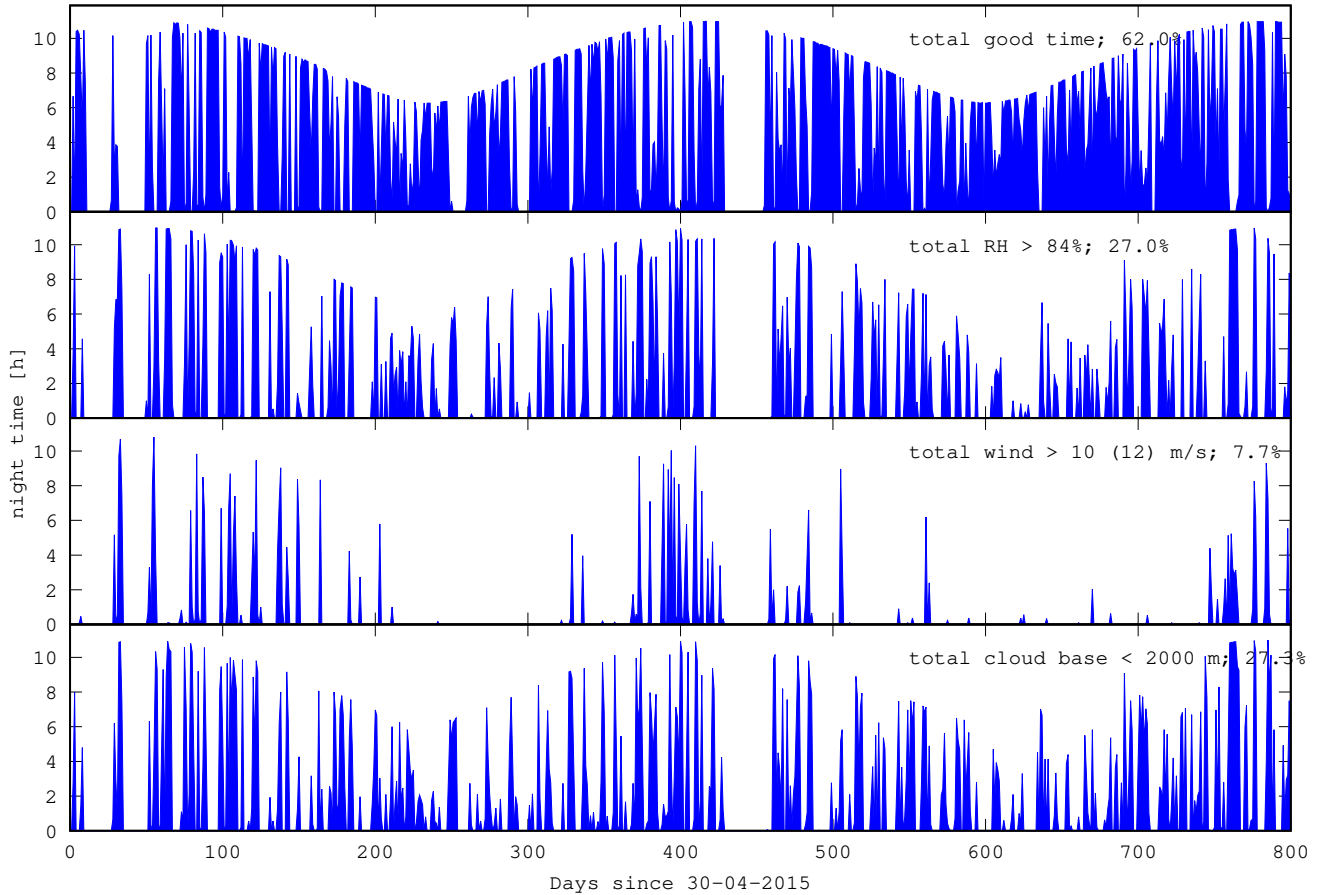}
\caption{Observing conditions analysis for SAAO during 800 days staring 30-04-2013. The top panel shows the amount of night hours per day with weather parameters within allowable limits. Dark hours are defined such that the Sun is lower than 18 degrees below the horizon. A total of 62.0\% of night time had favourable observing conditions. Gaps in the data that are the effect of system downtime or other abnormalities are not taken into account. The three lower panels show the contribution of relative humidity (RH), wind speed and cloud base into the non-observing time. Two wind speed parameters are used - average wind speed and maximum wind speed (the limits are 10 and 12 ms$^{-1}$, respectively). Precipitation is not analysed as it is a subset of the cloud contribution. Current observing limits related to ambient conditions are listed in Tab. \ref{tab:ObservingLimits}}.
\label{fig:weatherSAAO}
\end{figure*}

As described in the previous sections all three Solaris sites are equipped with weather monitoring equipment. This not only permits autonomous observing but  also allows us to analyze weather measurement data that is stored in the database. Figure \ref{fig:weatherSAAO} illustrates how different weather parameters influence the observability at SAAO. Relative humidity has the most significant contribution to the non-observing night time (27.0\%), followed by cloud cover (27.3\%) and wind speed (7.7\%). In practice the total count of night time hours with favourable weather conditions will be somewhat smaller due to several factors. One of them is the hysteresis (easing time) that is associated with each observing parameter independently and prevents the system from repeatedly closing and opening the  dome during unsure weather conditions. Secondly, the cloud base contribution is underestimated. Due to the operating principle of the cloud sensor, hight altitude clouds can be sometimes undetected which leads to worse quality data being collected.

\subsubsection{Observing limits}

Autonomous operation requires well defined observing limits to ensure that observations are only carried out during good weather conditions. The limits are, in general, more conservative than they would be for a human observer to maintain a larger safety margin in case of unexpected equipment failures. Not only ambient weather conditions are taken into account. Safe operation requires that parameters such as rack temperature, humidity, UPS battery capacity are monitored as well. Table \ref{tab:ObservingLimits} lists the most important parameters and their allowed values for Solaris-1. Most limits are identical throughout the sites. Cloud base is an exception due to different altitudes of the observatories that need to be accounted for.

\begin{deluxetable}{lcc}
\tabletypesize{\scriptsize}
\tablecaption{Example observing limits for Solaris-1. If a parameter value falls out of the allowable range it is considered to be in this state for the amount of time defined by the easing time value.}
\tablehead{\colhead{Variable} & \colhead{Limits} & \colhead{Easing time (min)}}
\startdata			
Temperature & $-1<T<30$ deg. C & 15 \\
Ambient RH & $RH<84$\%  & 15 \\
Minimum wind speed & $v_{min}<9$ ms$^{-1}$ & 25 \\
Average wind speed &  $v_{avg}<10$ ms$^{-1}$ & 25 \\
Maximum wind speed &  $v_{max}<12$ ms$^{-1}$ & 25 \\
Rain intensity & $i=0$ & 30 \\
Cloud base altitude & $h>3000$ m & 10 \\
Rack RH & $RH < 83$\%& 10 \\
Rack temperature & $0<T < 32$ deg. C& 10 \\
UPS battery capacity & $c > 80$\% & 10
\enddata
\label{tab:ObservingLimits}
\end{deluxetable}

\subsection{Observing workflow}

\begin{figure}[htb!]
\centering
\includegraphics[width=\columnwidth]{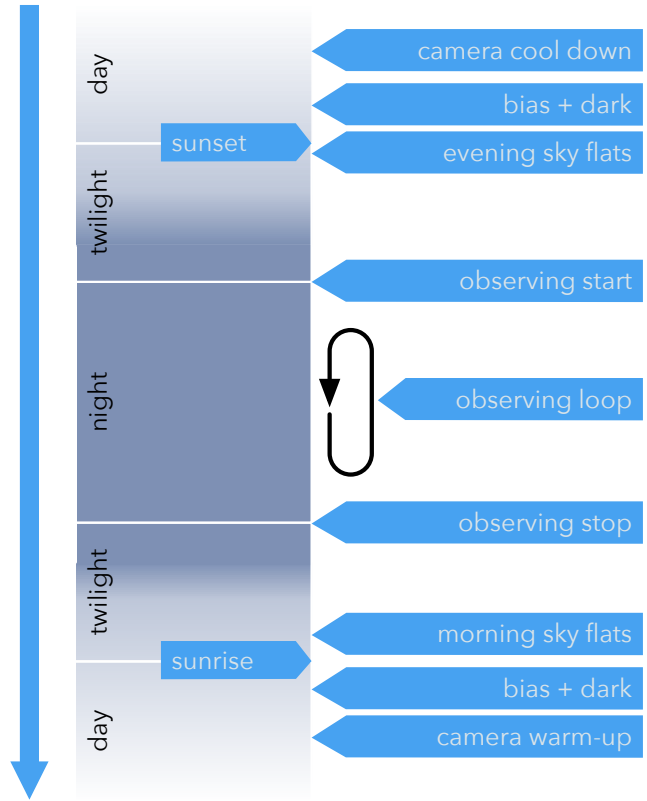}
\caption{Solaris high-level observing workflow. Stages in the diagram are described in text.}
\label{fig:Workflow}
\end{figure}

Observing workflow is an ordered list of steps that defines the behavior of the system during daytime, twilight and nighttime. An overview of the daily workflow is shown in Fig. \ref{fig:Workflow}. The system is in sleeping state during daytime. During this stage leftover data from the previous night is uploaded to the the servers in Poland. All equipment remains switched on. At a defined moment before sunset the system starts to prepare for observing. The photometric CCD camera is cooled down to its operating temperature of -70 deg. C. Sets of bias and dark frames are acquired. After the Sun sets and the light sensor's output drops below the defined threshold, the dome opens and flatfields are acquired. The order of filters and amount of flatfield frames to be acquired are defined in a configuration file. The exposure times are adjusted automatically to guarantee that the frames are properly exposed. If for a given filter the exposure time is too short (the shutter becomes evident in such case) the system delays the execution of the remaining flatfields. As the sky darkens, the remaining flatfields are acquired. After the flatfielding procedure is complete, the dome closes and the system waits for twilight to end. Once this happens, the observing queue is populated and the observing loop starts. The queue can be populated in two ways -- from a manually created observing program  in form of an xml file or from a cloud server. In the first case the user needs to make sure that the selected targets are observable in the order defined in the file. This mode is used when targets that are not in the global observing schedule need to be observed, usually for one night or part of a night. Normally, the observing queue is fed with targets from the cloud server. This service provides small chunks of the observing queue that are generated upon request for the specific site multiple times over the night. This scheduler makes sure that long-term project goals are accomplished in the global scale. Each element in the observing queue includes the name of the program, target identifier, field coordinates (that usually differ from the target coordinates), exposure time, filter band and a boolean value that decides whether exposure time should be automatically modified based on pixel counts from previous exposures. This is a very useful feature that helps in optimizing the exposure time and compensates for zenith distance and seeing changes throughout the night. Another functionality implemented in the system is astrometric frame solving and position correction. This feature is used to correct pointing errors and precisely position the field on the CCD frame, e.g. when one or more very bright stars need to positioned just outside the field of view. At all times all sensor data is read out processed and if necessary, the dome is closed an the observing queue paused. The observing queue is cleared at morning twilight and the dome closes. Morning flatfields are taken before sunrise following the same procedure as evening flatfields. Once this is done, sets of bias and dark frames are acquired. After that the system warms-up the CCD camera, enters sleeping state and remains in it during the day. All operations described above are executed asynchronously whenever possible. Multiple distributed security features on different levels overlook the safety of the system.

\subsection{Faults and major problems}
\label{sec:Faults}
We have experienced a number of equipment failures and problems. Probably the most notable ones concern the Andor CCD cameras. All four required servicing at least once at some point of the project. Considering the process of handling and shipping the camera from the observatory to the manufacturer and back, going through customs, etc., every failure of the camera meant a 2-3 month downtime in the operation of the observatory. Other hardware problems concerned the Vaisala weather transmitter (communication module replacement), Moxa serial card and fiber-ethernet converters. The PC internals in Solaris-3 required replacement due to problems with the motherboard. We have also experienced multiple UPS failures in Solaris-3, one of which involved a totally burnt unit. Luckily the security system isolated power from the equipment and nothing else had been damaged.

In January 2013 a severe bushfire passed through the Warrumbungle National Park threatening the Siding Spring Observatory. Solaris-3, as well as all other domes on the mouton survived the threat without damage. A few buildings and offices, however, have been destroyed. Since the service visit following the bushfire in early 2013, Solaris-3 has been operating for four years without the need for another maintenance visit.
In July 2013 a hail storm passed through the South African Astronomical Observatory. Two air-conditioning units (outdoor) required repair. 
%


\section{Scientific commissioning}
\label{sec:FirstResults}
In this section we present scientific results obtained with the Solaris telescopes during the commissioning phase. We demonstrate capabilities of individual telescopes and the network as a whole in delivering high quality scientific data. We focus mostly on photometry, but for selected cases we combine it with spectroscopy to present complete astrophysical models of eclipsing binaries.

\subsection{Photometric results}

Data reduction poses many challenges, especially if it is automated. Currently, a custom data reduction pipeline is being tested and verified. Part of the commissioning data presented in this paper has been reduced with this pipeline to demonstrate its possibilities. Unlike off-the shelf software packages, the dedicated pipeline is tailored to efficiently reduce data gathered by the Solaris telescopes and take full advantage of their capabilities such as shutter effects and camera nonlinearity modeling and SQL database interface (over 2M frames have been acquired).
The list of objects observed photometrically comprises exoplanetary transits and eclipsing binaries of various types and is presented in Tab. \ref{tab:campaign}. The range of $V$ magnitudes of the observed objects spans 8.99 to 14.40.

\begin{deluxetable*}{llrll}
\tablecaption{Targets observed with the Solaris network during commissioning and testing phase.}
\tablehead{\colhead{Object ID} & \colhead{J2000 Coordinates} & \colhead{V mag} & \colhead{Telescope} & \colhead{Comments}}
\startdata
Wasp-4b 			& \ra{23}{34}{15}{06}  \de{--42}{03}{41}{10}	& 12.5   & SLR1 &  exoplanet \\
Wasp-64b 		& \ra{06}{44}{27}{61}  \de{--32}{51}{30}{25} 	& 12.29 & SLR4 &  exoplanet \\
Wasp-98b 		& \ra{03}{53}{42}{91}  \de{--34}{19}{41}{50}	& 13.00 & SLR3 &  exoplanet \\
PG1336-0118		& \ra{13}{38}{48}{15}  \de{--02}{01}{49}{10}	& 13.30 & SLR2 & close eclipsing binary with pulsating subdwarf component\\
RR Cae			& \ra{04}{21}{05}{56}  \de{--48}{39}{07}{02}	& 14.40 & SLR3 & white and red dwarf eclipsing binary with mass transfer \\
KZ Hya			& \ra{10}{50}{54}{08}  \de{--25}{21}{14}{71}	& 10.06 & SLR3 & short-period high amplitude pulsating variable \\
SOL-0023			& \multicolumn{2}{c}{undisclosed}				& SLR1	  & eclipsing binary \\					
J024946-3825.6	& \ra{02}{49}{45}{90}  \de{--38}{25}{36}{00}	& 13.30 & SLR1, SLR3, SLR4 & eclipsing binary, SOL-0132
\enddata
\label{tab:campaign}
\end{deluxetable*}

\subsubsection{Exoplanet transits}
In this section we present exoplanetary transits obtained using Solaris-1, Solaris-3 and Solaris-4 telescopes, covering all three sites of the network. Presented data sets have been reduced using the AstroImageJ (AiJ) data reduction package \citep{Collins2016}. AstroImageJ provides a very intuitive and user friendly graphical interface that greatly simplifies photometric data reduction. Its interactive capabilities are especially convenient for reducing data gathered in short observing runs. Transit fitting was done using the EXOFAST code\citep{Eastman2013, Eastman2010, Wright2014} - and IDL package for transit modeling that implements routines allowing simultaneous or separate light curve transit and radial velocity fitting. The biggest strength of EXOFAST, however, is its ability to characterize the parameter uncertainties using a differential evolution Markov chain Monte Carlo method. Exoplanet transits were selected using the Exoplanet Transit Database \citep{Brat2010} that also provided ephemerides and visibility information.

\begin{figure}[htbp]
\centering
\includegraphics[width=\columnwidth]{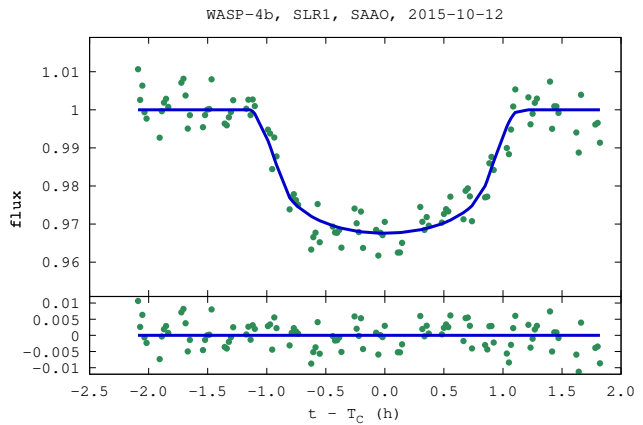}
\caption{Normalized and light curve data for WASP-4b with best fit transit model overlaid.}
\label{fig:wasp-4}
\end{figure}

\paragraph{WASP-4b}
Wasp-4b is a hot Jupiter transiting a $V = 12.5$ mag star discovered by the SuperWASP-South observatory and CORALIE collaboration. The transit was recorded with the Solaris-1 telescope in SAAO on the evening of October 11th, 2015 in the $V$ band. The exposure time was fixed at 59 seconds. A master bias frame wes subtracted from our raw science images, then a median twilight flatfield frame was used to remove image inhomogenities. Aperture photometry was used with fixed star apertures and sky background annulus. Apertures were automatically recentered using the center-of-light method \citep{Howell2006} implemented in AiJ. Occasionally the aperture matching and recentering algorithm required manual intervention. The host star is reported to be a G7V main sequence star with $T_{\textrm{eff}} = 5500\pm150$K, log$g = 4.3\pm 0.2$, [M/H] = $0.0\pm0.2$ \citep{Wilson2008}. The remaining priors used as input for EXOFAST include period, inclination, RV semi-amplitude and semimajor axis - all taken from the aforecited discovery paper. The time of transit prior has been estimated using AiJ, a circular orbit was assumed. The period value was kept fixed since only one transit was observed. Photometric data, obtained model and residuals are plotted in Fig. \ref{fig:wasp-4}. The system's parameters along with errors are listed in Tab. \ref{tab:wasp-4}.  The fit's RMS is 3.6 mmag. \cite{Wilson2008} published RMS scatter values of 15.3, 2.7 and 1.8 mmag for data obtained with WASP-S, the 2.0-m Faulkes Telescope South and 1.2-m Euler Telescope, respectively. With a fully developed data reduction pipeline we expect even better photometric precision. 

\begin{deluxetable}{lcc}
\tablecaption{Median values and 68\% confidence interval for Wasp-4b obtained with EXOFAST.}
\tablehead{\colhead{~~~Parameter} & \colhead{Units} & \colhead{Value}}
\startdata
\sidehead{Stellar Parameters:}
                           ~~~$M_{*}$\dotfill &Mass (\msun)\dotfill & $0.962_{-0.069}^{+0.073}$\\
                         ~~~$R_{*}$\dotfill &Radius (\rsun)\dotfill & $0.970_{-0.027}^{+0.030}$\\
                     ~~~$L_{*}$\dotfill &Luminosity (\lsun)\dotfill & $0.78_{-0.11}^{+0.12}$\\
                         ~~~$\rho_*$\dotfill &Density (cgs)\dotfill & $1.485_{-0.068}^{+0.069}$\\
              ~~~$\log(g_*)$\dotfill &Surface gravity (cgs)\dotfill & $4.447\pm0.017$\\
              ~~~$\teff$\dotfill &Effective temperature (K)\dotfill & $5510\pm150$\\
                              ~~~$\feh$\dotfill &Metalicity\dotfill & $-0.00\pm0.20$\\
\sidehead{Planetary Parameters:}
                              ~~~$P$\dotfill &Period (days)\dotfill & $1.3382281\pm0.0000030$\\
                       ~~~$a$\dotfill &Semi-major axis (AU)\dotfill & $0.02346_{-0.00057}^{+0.00058}$\\
                           ~~~$R_{P}$\dotfill &Radius (\rj)\dotfill & $1.515_{-0.052}^{+0.057}$\\
           ~~~$T_{eq}$\dotfill &Equilibrium Temperature (K)\dotfill & $1707\pm47$\\
               ~~~$\fave$\dotfill &Incident flux (\fluxcgs)\dotfill & $1.93_{-0.20}^{+0.22}$\\
\sidehead{Primary Transit Parameters:}
                ~~~$T_C$\dotfill &Time of transit (\bjdtdb)\dotfill & $2457307.34753\pm0.00023$\\
~~~$R_{P}/R_{*}$\dotfill &Radius of planet in stellar radii\dotfill & $0.1605\pm0.0033$\\
     ~~~$a/R_{*}$\dotfill &Semi-major axis in stellar radii\dotfill & $5.198_{-0.080}^{+0.079}$\\
              ~~~$u_1$\dotfill &linear limb-darkening coeff\dotfill & $0.520_{-0.059}^{+0.063}$\\
           ~~~$u_2$\dotfill &quadratic limb-darkening coeff\dotfill & $0.222_{-0.056}^{+0.055}$\\
                      ~~~$i$\dotfill &Inclination (degrees)\dotfill & $88.50_{-1.0}^{+0.93}$\\
                           ~~~$b$\dotfill &Impact Parameter\dotfill & $0.136_{-0.084}^{+0.090}$\\
                         ~~~$\delta$\dotfill &Transit depth\dotfill & $0.0258\pm0.0011$\\
                ~~~$T_{FWHM}$\dotfill &FWHM duration (days)\dotfill & $0.0814_{-0.0014}^{+0.0013}$\\
          ~~~$\tau$\dotfill &Ingress/egress duration (days)\dotfill & $0.01355_{-0.00040}^{+0.00049}$\\
                 ~~~$T_{14}$\dotfill &Total duration (days)\dotfill & $0.0950\pm0.0014$\\
      ~~~$P_{T}$\dotfill &A priori non-grazing transit prob\dotfill & $0.1615_{-0.0025}^{+0.0026}$\\
                ~~~$P_{T,G}$\dotfill &A priori transit prob\dotfill & $0.2233_{-0.0034}^{+0.0035}$\\
                            ~~~$F_0$\dotfill &Baseline flux\dotfill & $1.4802\pm0.0012$\\
\sidehead{Secondary Eclipse Parameters:}
              ~~~$T_{S}$\dotfill &Time of eclipse (\bjdtdb)\dotfill & $2457308.01664_{-0.00023}^{+0.00024}$
\enddata
\label{tab:wasp-4}
\end{deluxetable}

\begin{figure}[htb!]
\centering
\includegraphics[width=\columnwidth]{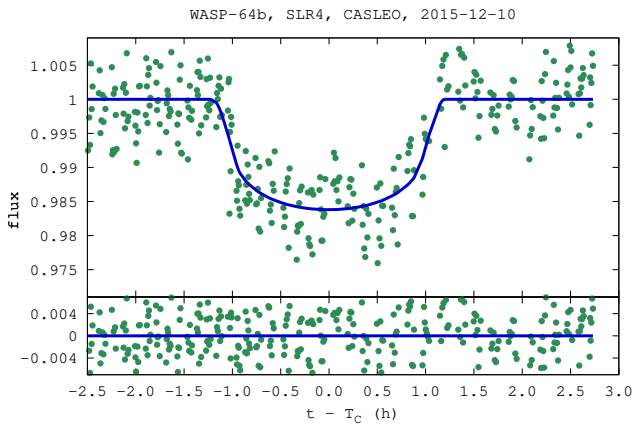}
\caption{Normalized and light curve data for Wasp-64b with best fit transit model overlaid.}
\label{fig:wasp-64}
\end{figure}

\paragraph{WASP-64b}
The discovery of Wasp-64b has been reported by \cite{Gillon2013}. It is a $1.271\rj$  and $1.271\mj$ giant planet in a short ($a=0.02648$ AU, $P = 1.5732918$ d) orbit around a $V = 12.3$ mag G7-type dwarf. The transit has been recorded using the Solaris-4 telescope in CASLEO on December 10th, 2015. 313 frames were acquired in the $V$-band with varying auto-adjusted exposure time between 20 and 40 seconds. Priors for EXOFAST were taken from the aforecited discovery paper. Photometric data, obtained model and residuals are plotted in Fig. \ref{fig:wasp-64}. The system's parameters along with errors are listed in Tab. \ref{tab:wasp-64}. The fit's RMS is 2.6 mmag. 

\begin{deluxetable}{lcc}
\tablecaption{Median values and 68\% confidence interval for Wasp-64b obtained with EXOFAST.}
\tablehead{\colhead{~~~Parameter} & \colhead{Units} & \colhead{Value}}
\startdata
\sidehead{Stellar Parameters:}
                           ~~~$M_{*}$\dotfill &Mass (\msun)\dotfill & $0.959_{-0.061}^{+0.063}$\\
                         ~~~$R_{*}$\dotfill &Radius (\rsun)\dotfill & $1.039_{-0.044}^{+0.045}$\\
                     ~~~$L_{*}$\dotfill &Luminosity (\lsun)\dotfill & $0.89_{-0.13}^{+0.15}$\\
                         ~~~$\rho_*$\dotfill &Density (cgs)\dotfill & $1.21_{-0.12}^{+0.13}$\\
              ~~~$\log(g_*)$\dotfill &Surface gravity (cgs)\dotfill & $4.387\pm0.030$\\
              ~~~$\teff$\dotfill &Effective temperature (K)\dotfill & $5500\pm150$\\
                              ~~~$\feh$\dotfill &Metalicity\dotfill & $-0.08\pm0.11$\\
\sidehead{Planetary Parameters:}
                              ~~~$P$\dotfill &Period (days)\dotfill & $1.5732917\pm0.0000015$\\
                       ~~~$a$\dotfill &Semi-major axis (AU)\dotfill & $0.02610\pm0.00056$\\
                           ~~~$R_{P}$\dotfill &Radius (\rj)\dotfill & $1.15_{-0.15}^{+0.14}$\\
           ~~~$T_{eq}$\dotfill &Equilibrium Temperature (K)\dotfill & $1672_{-53}^{+55}$\\
               ~~~$\fave$\dotfill &Incident flux (\fluxcgs)\dotfill & $1.77_{-0.21}^{+0.24}$\\
\sidehead{Primary Transit Parameters:}
                ~~~$T_C$\dotfill &Time of transit (\bjdtdb)\dotfill & $2457366.71230_{-0.00100}^{+0.00095}$\\
~~~$R_{P}/R_{*}$\dotfill &Radius of planet in stellar radii\dotfill & $0.114_{-0.015}^{+0.013}$\\
     ~~~$a/R_{*}$\dotfill &Semi-major axis in stellar radii\dotfill & $5.40_{-0.18}^{+0.19}$\\
              ~~~$u_1$\dotfill &linear limb-darkening coeff\dotfill & $0.511_{-0.062}^{+0.063}$\\
           ~~~$u_2$\dotfill &quadratic limb-darkening coeff\dotfill & $0.226_{-0.055}^{+0.053}$\\
                      ~~~$i$\dotfill &Inclination (degrees)\dotfill & $86.569\pm0.097$\\
                           ~~~$b$\dotfill &Impact Parameter\dotfill & $0.323_{-0.014}^{+0.015}$\\
                         ~~~$\delta$\dotfill &Transit depth\dotfill & $0.0129_{-0.0031}^{+0.0030}$\\
                ~~~$T_{FWHM}$\dotfill &FWHM duration (days)\dotfill & $0.0882_{-0.0034}^{+0.0036}$\\
          ~~~$\tau$\dotfill &Ingress/egress duration (days)\dotfill & $0.0113_{-0.0015}^{+0.0013}$\\
                 ~~~$T_{14}$\dotfill &Total duration (days)\dotfill & $0.0995\pm0.0040$\\
      ~~~$P_{T}$\dotfill &A priori non-grazing transit prob\dotfill & $0.1642_{-0.0061}^{+0.0065}$\\
                ~~~$P_{T,G}$\dotfill &A priori transit prob\dotfill & $0.2060\pm0.0075$\\
                            ~~~$F_0$\dotfill &Baseline flux\dotfill & $0.20836_{-0.00044}^{+0.00041}$\\
\sidehead{Secondary Eclipse Parameters:}
              ~~~$T_{S}$\dotfill &Time of eclipse (\bjdtdb)\dotfill & $2457367.49895_{-0.00100}^{+0.00095}$
\enddata
\label{tab:wasp-64}
\end{deluxetable}

\paragraph{Wasp-98b}

\begin{figure}[htb!]
\centering
\includegraphics[width=\columnwidth]{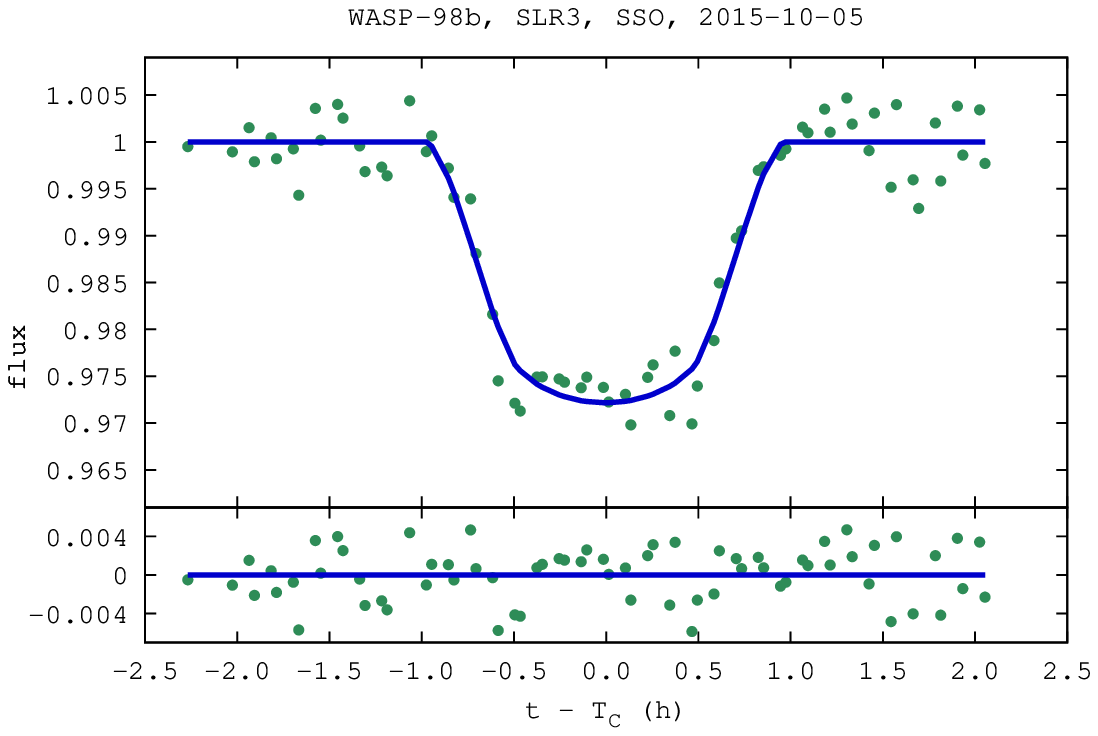}
\caption{Normalized and light curve data for Wasp-98b with best fit transit model overlaid.}
\label{fig:wasp-98}
\end{figure}

Wasp-98b was discovered by \cite{Hellier2014}. It is a $0.83\mj$ hot Jupier orbiting a G7-type star. Photometric data, obtained model and residuals are plotted in Fig. \ref{fig:wasp-98}. The system's parameters along with errors are listed in Tab. \ref{tab:wasp-98}.  The fit's RMS is  2.9 mmag. 

\begin{deluxetable}{lcc}
\tablecaption{Median values and 68\% confidence interval for Wasp-98b obtained with EXOFAST.}
\tablehead{\colhead{~~~Parameter} & \colhead{Units} & \colhead{Value}}
\startdata
\sidehead{Stellar Parameters:}
                           ~~~$M_{*}$\dotfill &Mass (\msun)\dotfill & $0.840_{-0.050}^{+0.055}$\\
                         ~~~$R_{*}$\dotfill &Radius (\rsun)\dotfill & $0.754\pm0.020$\\
                     ~~~$L_{*}$\dotfill &Luminosity (\lsun)\dotfill & $0.488_{-0.061}^{+0.069}$\\
                         ~~~$\rho_*$\dotfill &Density (cgs)\dotfill & $2.77_{-0.14}^{+0.15}$\\
              ~~~$\log(g_*)$\dotfill &Surface gravity (cgs)\dotfill & $4.609\pm0.018$\\
              ~~~$\teff$\dotfill &Effective temperature (K)\dotfill & $5560\pm140$\\
                              ~~~$\feh$\dotfill &Metalicity\dotfill & $-0.601_{-0.10}^{+0.098}$\\
\sidehead{Planetary Parameters:}
                              ~~~$P$\dotfill &Period (days)\dotfill & $2.9626401_{-0.0000014}^{+0.0000013}$\\
                       ~~~$a$\dotfill &Semi-major axis (AU)\dotfill & $0.03808_{-0.00078}^{+0.00082}$\\
                           ~~~$R_{P}$\dotfill &Radius (\rj)\dotfill & $1.199\pm0.032$\\
           ~~~$T_{eq}$\dotfill &Equilibrium Temperature (K)\dotfill & $1193\pm31$\\
               ~~~$\fave$\dotfill &Incident flux (\fluxcgs)\dotfill & $0.460_{-0.046}^{+0.050}$\\
\sidehead{Primary Transit Parameters:}
                ~~~$T_C$\dotfill &Time of transit (\bjdtdb)\dotfill & $2457305.13972\pm0.00049$\\
~~~$R_{P}/R_{*}$\dotfill &Radius of planet in stellar radii\dotfill & $0.163476_{-0.000100}^{+0.000097}$\\
     ~~~$a/R_{*}$\dotfill &Semi-major axis in stellar radii\dotfill & $10.87\pm0.19$\\
              ~~~$u_1$\dotfill &linear limb-darkening coeff\dotfill & $0.410_{-0.056}^{+0.060}$\\
           ~~~$u_2$\dotfill &quadratic limb-darkening coeff\dotfill & $0.266_{-0.054}^{+0.052}$\\
                      ~~~$i$\dotfill &Inclination (degrees)\dotfill & $86.324_{-0.091}^{+0.092}$\\
                           ~~~$b$\dotfill &Impact Parameter\dotfill & $0.697_{-0.012}^{+0.011}$\\
                         ~~~$\delta$\dotfill &Transit depth\dotfill & $0.026724_{-0.000033}^{+0.000032}$\\
                ~~~$T_{FWHM}$\dotfill &FWHM duration (days)\dotfill & $0.0607\pm0.0015$\\
          ~~~$\tau$\dotfill &Ingress/egress duration (days)\dotfill & $0.02043_{-0.00054}^{+0.00055}$\\
                 ~~~$T_{14}$\dotfill &Total duration (days)\dotfill & $0.0811\pm0.0016$\\
      ~~~$P_{T}$\dotfill &A priori non-grazing transit prob\dotfill & $0.0769_{-0.0013}^{+0.0014}$\\
                ~~~$P_{T,G}$\dotfill &A priori transit prob\dotfill & $0.1070\pm0.0019$\\
                            ~~~$F_0$\dotfill &Baseline flux\dotfill & $1.18441\pm0.00054$\\
\sidehead{Secondary Eclipse Parameters:}
              ~~~$T_{S}$\dotfill &Time of eclipse (\bjdtdb)\dotfill & $2457306.62104\pm0.00049$
\enddata
\label{tab:wasp-98}
\end{deluxetable}	

\subsubsection{Timing targets}

The main goal of Project Solaris is to monitor and detect timing variations of eclipsing binaries. In this section we present photometric measurements of eclipsing and pulsating binaries and demonstrate the achievable photometric precision. 
\paragraph{KZ Hya}

\begin{figure}[htb!]
\center
\includegraphics[width=\columnwidth]{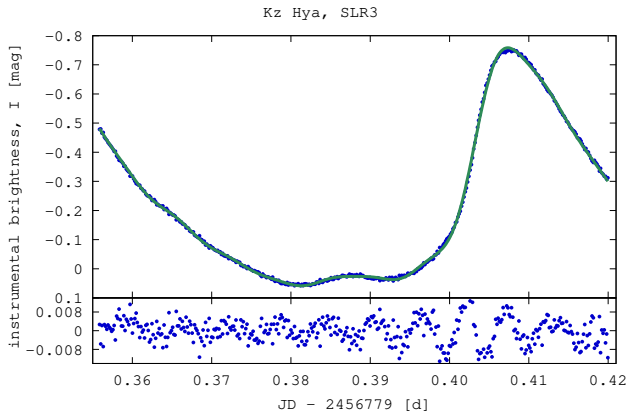}
\caption{KZ Hya, V-band data for one epoch with (top panel). Residuals after fitting nine frequencies from \cite{Kim2007}.}
\label{fig:KZHya}
\end{figure}

KZ Hya or HD94033 is a SX Phoenicis type pulsating variable. Sx Phe stars are metal-poor Type II population variables that pulsate with periods between 1.0 and 1.75 h. 
\citep{McNamara1995}. These pulsations can act as a convenient beacon and reveal additional information about the pulsator and its possible companion. Indeed, \cite{Kim2007} show that an 24.77 year eccentric ($e = 0.25$) orbit is visible in the O--C data. Assuming that the primary has a mass of 0.9\msun, the minimum mass of the unseen companion is  0.66\msun$\sin i$. This target has been selected for a follow-up campaign. At this time we present an example V-band light curve of a single epoch obtained with the Solaris-3 telescope -- Fig. \ref{fig:KZHya}. Frequencies from \cite{Kim2007} have been fitted to the data. Residuals have an rms of 5 mmag, but the short term precision is 1 mmag. KZ Hya will be investigated in much more detail in an upcoming publication.

\paragraph{RR Cae}

\begin{figure}[htb!]
\centering
\includegraphics[width=\columnwidth]{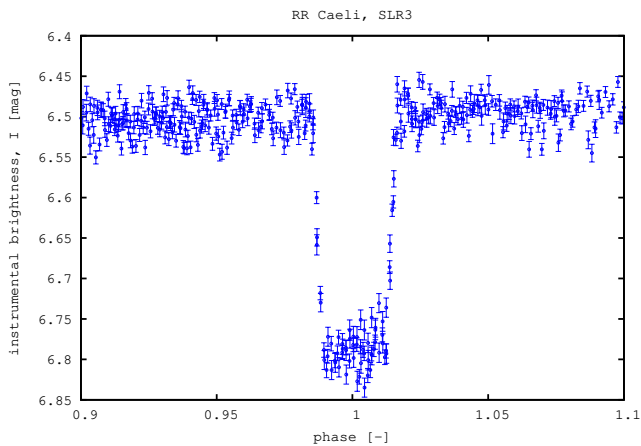}
\caption{RR Cae primary eclipse recorded in the I band. Data has been phased with $P=0.303704$ d and spans 62 days from March 11th, 2015.}
\label{fig:RRCae}
\end{figure}

RR Cae is a $V=14.4$ mag short period dwarf-M-dwarf eclipsing binary. \cite{Maxted2007} have shown that no meaningful O--C variations are present in the data spanning 10 years. A more recent study by \cite{Qian2012}, however, reveals a periodic signal in O--C measurements. After ruling out several other effects that might have been the reason for the 11.9 year variations, the authors conclude that a circumbinary planet is responsible for the detected signal. RR Cae is therefore an interesting target that is present in our long-term timing campaign. An example light curve around the primary eclipse is shown in Fig. \ref{fig:RRCae}.

\paragraph{SOL-0023}

\begin{figure}[htb!]
\centering
\includegraphics[width=\columnwidth]{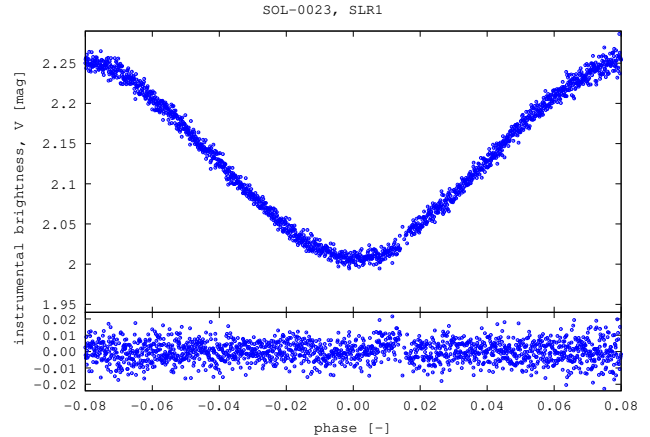}
\caption{SOL-0023 phased V-band light curve around the primary eclipse used for eclipse timing analysis.}
\label{fig:SOL_0023}
\end{figure}

SOL-0023 is a bright, $\sim 2.5$ d eclipsing binary that is a very promising target in our eclipse timing campaign. Several mmag precision is achievable regularly with the Solaris telescopes. A sample light curve centered around the primary eclipse is shown in Fig. \ref{fig:SOL_0023}.

\subsubsection{Eclipsing binary with a pulsator}

\paragraph{PG 1336-018}

\begin{figure}[htb!]
\centering
\includegraphics[width=\columnwidth]{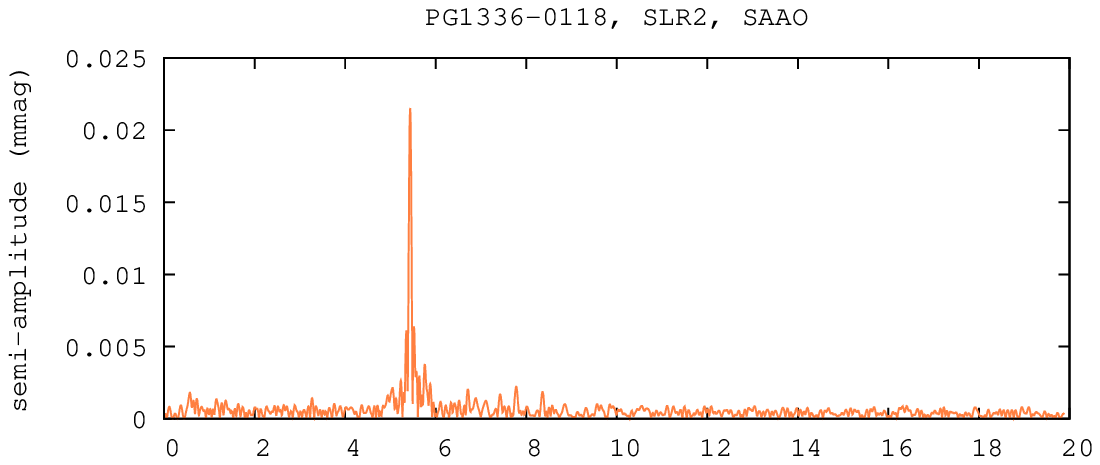}
\includegraphics[width=\columnwidth]{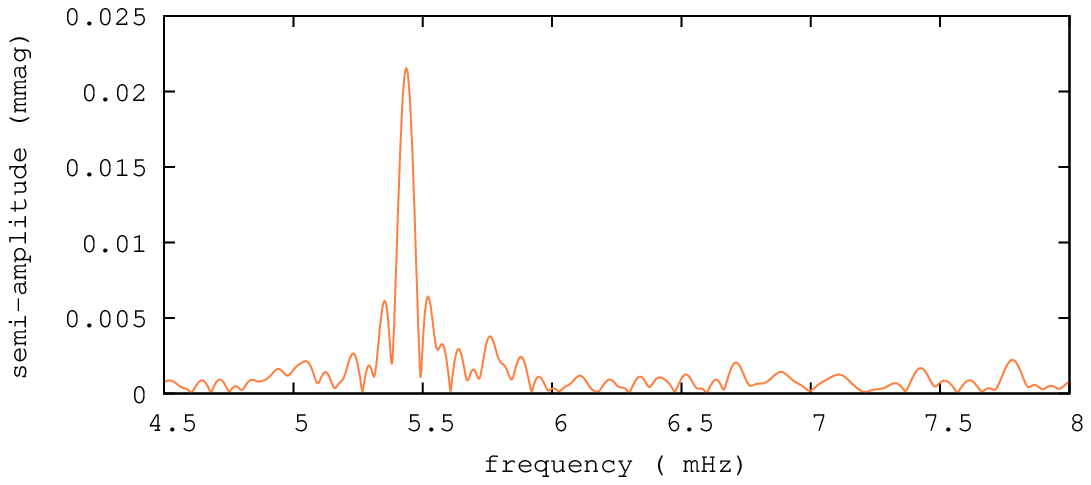}
\caption{PG1336-0118 frequency analysis. Top panel shows amplitude spectrum in the entire computed range. Bottom panel shows more detail in the 5.5 mHz frequency range.}
\label{fig:24hFourier}
\end{figure}

\begin{figure}[htb!]
\centering
\includegraphics[width=\columnwidth]{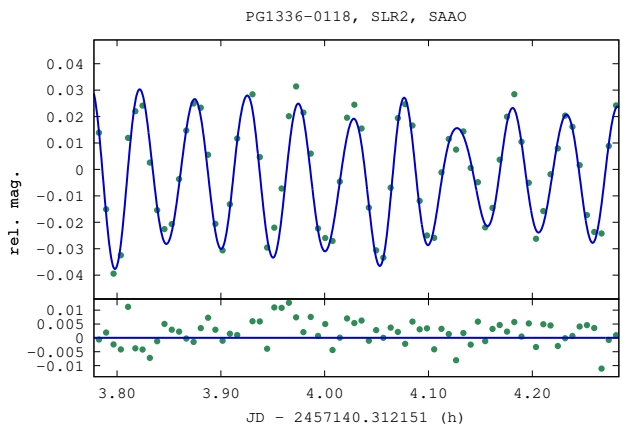}
\caption{Part of residuals of PG1336-018 and fitted oscillation model with four identified frequencies.}
\label{fig:24hperiodogram}
\end{figure}

\begin{figure*}[htb!]
\centering
\includegraphics[width=\textwidth]{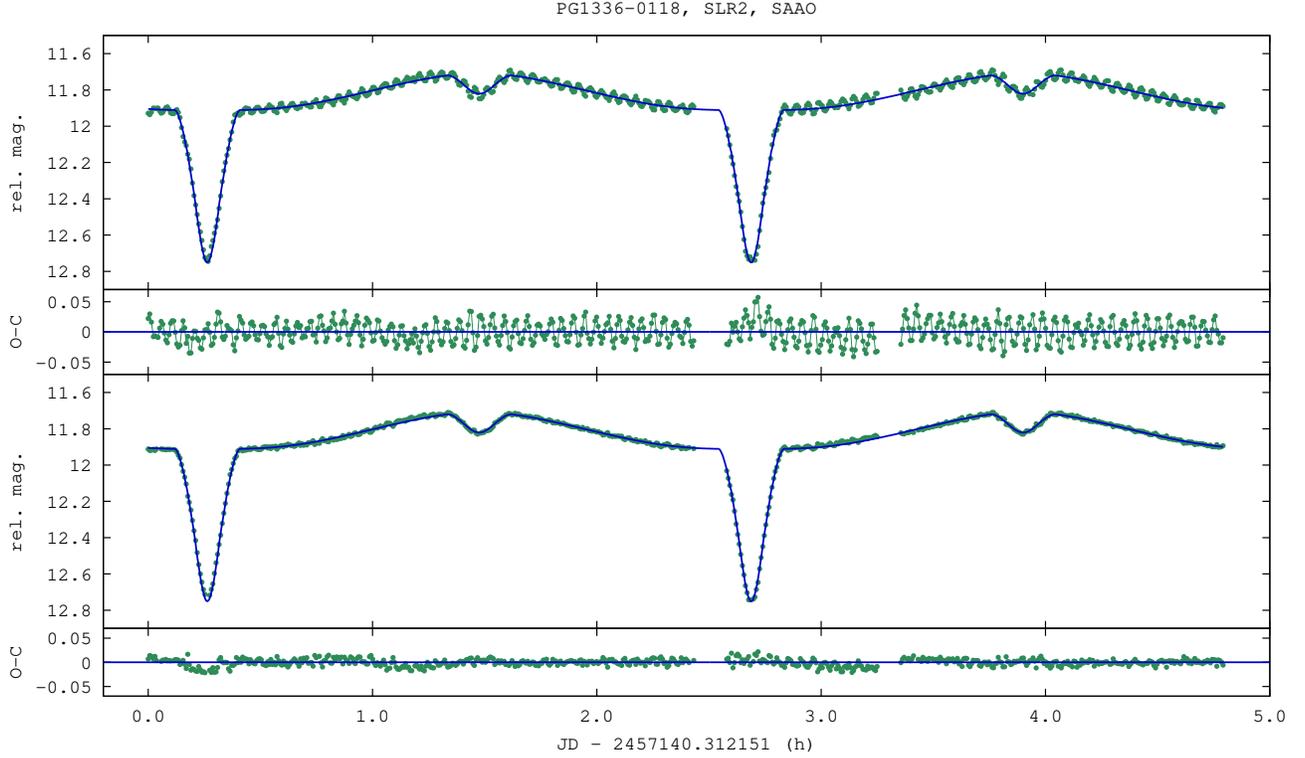}
\caption{PG1336-0118 Solaris-2 light curve and models. The first panel shows the observed magnitude and the initial eclipsing binary model obtained with PHOEBE. The second panel shows residuals of the above fit. Points have been connected with lines to better visualize pulsations. The third panel shows the data from the first panel with pulsations removed based on the frequencies, amplitudes and phases identified during the analysis of the residuals. An updated model is over-plotted on the data. Residuals of this model are shown in the last panel. Visible trends are caused by imperfect modeling of the reflection effects.}
\label{fig:24h}
\end{figure*}

PG 1336-018 (NY Vir) is a $V=13.3$ mag eclipsing B-type subdwarf (sdB). With $P=0.1010174$ d it is one of the shortest-period eclipsing binary known. Additionally, the system shows pulsations of the type found in sdB pulsators. \cite{Hu2007} present a detailed evolutionary study of the system. Its history is complex: in the past the system underwent the common-envelope stage with a companion of unknown type; after mass transfer, the system evolved into what we observe now - a binary pulsating star with a M-type dwarf companion. \cite{Kilkenny1998} studied this binary initially and identified pulsations with periods 184 s and 141 s and semi-amplitudes 10 and 5 mmag, respectively. Data was collected with the University of Cape Town (UCT) 1-m telescope using a CCD and photometer. The authors expected additional pulsation frequencies with semi-amplitudes below 3 mmag. These have been identified in the follow-up study \citep{Kilkenny2003} that was based on a multi-site (Whole Earth Telescope) observing campaign. Authors detected and identified 28 frequencies down to the semi-amplitude of 0.5 mmag. The most recent study by \cite{Vuckovic2007} presents a complete astrophysical model of the system based on photometric and spectroscopic data obtained with the VLT (UVES and ULTRACAM). PG 1336-016 was observed with the Solaris-2 telescope on the night of April 24th 2015. 650 20-second exposures were acquired without any filter. The decision not to use filters was made in order to increase photon counts and keep the exposure time at 20 s to obtain good temporal resolution and satisfactory SNR of the target and comparison stars. Frames were reduced with AiJ using a similar approach as for exoplanetary transits. Relative fluxes obtained with AiJ were loaded into \textsc{phoebe} \citep{Prsa2005} and \textsc{jktebop} \citep{sou04a,sou04b}, where an initial model was fitted. Priors for physical parameters were adopted from \cite{Vuckovic2007}. Once a satisfactory fit was obtained, residuals were analyzed with the Period04 software package \citep{Lenz2005}. Several frequencies have been identified in the data (Fig. \ref{fig:24hFourier}). Monte Carlo analysis was performed to obtain uncertainties of the frequencies and corresponding semiaamplitudes. Table \ref{tab:24hfrequencies} lists four frequencies with highest semiamplitudes that have been used to remove oscillations from the data set. Figure \ref{fig:24hperiodogram} illustrates a 30-minute long set of data points with the fitted oscillation model. Once cleaned, a second iteration fit was done with \textsc{phoebe} and \textsc{jktebop} to obtain a final set of parameters. Uncertainties of parameters have been computed using \textsc{jktabsdim}. The resulting light curve and model are shown in Fig. \ref{fig:24h}.
This target has been chosen to demonstrate the capabilities of the network to acquire high cadence for asteroseismology. The obtained astrophysical model, constructed primarily based on new data with priors from the literature, is in agreement with what can be found in previous works.

\begin{deluxetable*}{cccccc}
\tablewidth{0pt}
\tabletypesize{\scriptsize}
\tablecaption{Four frequencies identified in the residuals' spectrum of PG1336-018. Obtained values are the result of 1000 Monte Carlo simulations computed using Period04.}
\tablehead{\colhead{F\#} & \colhead{$f$ ($\mu$ Hz)} & \colhead{$1/f (s)$} & \colhead{$\sigma f$ ($\mu$ Hz)} & \colhead{semiampl. (mmag)} &  \colhead{$\sigma$ semiampl. (mmag)} }
\startdata
F1	& 5436.2	 	& 183.95	& 0.6 		& 21.7 		& 0.4\\
F2	& 5540		& 180.50		& 6		& 2.1		& 0.4\\
F3	& 5894		& 169.66		& 6		& 2.1		& 0.4\\
F4	& 6719		& 148.83		& 5  		& 2.1 	& 0.4
\enddata
\label{tab:24hfrequencies}
\end{deluxetable*}

\begin{deluxetable}{rlrr}
\tablewidth{0pt}

\tablecaption{Solutions for eclipsing binaries. Formal errors are noted directly under the parameter value.}
\tablehead{\colhead{Parameter}	&	\colhead{Unit}	&	\colhead{PG1336-018}	&	\colhead{J024946-3825.6}}
\startdata	
	$T_0$	&	(JD)	&	2457140.323152	&	2457353.37820 	\\ 		 \vspace{0.1cm}
		&		&	0.000008	&		0.00004 \\		
	$P$	&	(d)	&	0.101004	&	0.463220	\\		 \vspace{0.1cm}
		&		&	 0.000012	&	0.000022	\\		
	$K_1$	&	(km s$^{-1}$) 	&	78.6	&	124.3	\\		 \vspace{0.1cm}
		&		&	0.6	&	2.1	\\		
	$K_2$	&	(km s$^{-1}$) 	&	300	&	162.3	\\		 \vspace{0.1cm}
		&		&	2	&	2.6	\\		
	$e$	&		&	0	&	0	\\		 \vspace{0.1cm}
		&		&	-	&	-	\\		
	$i$	&	(deg.)	&	77.88	&	81.0	\\		 \vspace{0.1cm}
		&		&	0.27	&	1.0	\\		
	$a$	&	(R$_\sun$)	&	 0.461	&	2.66	\\		 \vspace{0.1cm}
		&		&	 0.006 	&	0.03	\\		
	$\omega$	&	(\arcdeg)	&	-	&	-	\\		 \vspace{0.1cm}
		&		&	-	&	-	\\		
	$v_\gamma$	&	(km s$^{-1}$) 	&	-	&	21.4	\\		 \vspace{0.1cm}
		&		&	-	&	0.5	\\		
	RMS RV1	&	(km s$^{-1}$) 	&	-	&	2.2	\\ 		 \vspace{0.1cm}
	RMS RV2	&	(km s$^{-1}$) 	&	-	&	1.4	\\		
	$T_1$	&	(K)	&	32850	&	4100	\\		 \vspace{0.1cm}
		&		&	-	&	350	\\		
	$T_2$	&	(K)	&	3100	&	3475	\\		 \vspace{0.1cm}
	~	&		&	-	&	350	\\		
	$M_1$	&	(M$_\sun$)	&	0.468	&	0.664	\\		 \vspace{0.1cm}
		&		&	0.008	&	0.025	\\		
	$M_2$	&	(M$_\sun$)	&	0.123	&	0.509	\\		 \vspace{0.1cm}
		&		&	0.002	&	0.019	\\		
	$R_1$	&	(R$_\sun$)	&	0.1448	&	0.590	\\		 \vspace{0.1cm}
		&		&	0.0021	&	0.027	\\		
	$R_2$	&	(R$_\sun$)	&	0.1543	&	0.518	\\		 \vspace{0.1cm}
		&		&	0.0022	&	0.027	\\		
	log $g_1$	&	(cm s$^{-1}$)	&	5.787	&	4.72	\\		 \vspace{0.1cm}
		&		&	0.012	&	0.04	\\		
	log $g_2$	&	(cm s$^{-1}$)	&	5.150	&	4.72	\\		 \vspace{0.1cm}
		&		&	 0.012	&	0.04	\\		
	$v_{synchr, 1}$	&	(km s$^{-1}$) 	&	72.5	&	64	\\		 \vspace{0.1cm}
		&		&	1.0	&	3	\\		
	$v_{synchr, 2}$	&	(km s$^{-1}$) 	&	 77.3	&	56.6	\\		 \vspace{0.1cm}
		&		&	1.1	&	3.0	\\		
	log $L_1$	&	(L$_\sun$)	&	1.343	&	-1.05	\\		 \vspace{0.1cm}
		&		&	0.015	&	 0.15	\\		
	log $L_2$	&	(L$_\sun$)	&	-2.70	&	-1.45	\\		 \vspace{0.1cm}
		&		&	0.09	&	0.18	\\		
	$M_{bol,1}$	&	(mag)	&	1.392	&	7.4	\\		 \vspace{0.1cm}
		&		&	0.038	&	0.4	\\		
	$M_{bol,2}$	&	(mag)	&	11.51	&	 8.4	\\		 \vspace{0.1cm}
		&		&	 0.23	&	0.5	\\		

\enddata
\label{tab:ModelsSolutions}
\end{deluxetable}

\subsubsection{Complete model - J024946-3825.6}

J024946-3825.6 or SOL-0132, also known as 1RXS\footnote{1-st ROSAT X Survey.} J024946.0-382540, is a $V=11.7$ mag eclipsing detached binary from the ASAS catalog with an orbital period of 0.46322 days. It has been observed photometrically with SLR1, SLR3 and SLR4 telescopes. Spectra were obtained with the SMARTS 1.5m telescope and the Chiron spectrograph at the Cerro Tololo Inter-American Observatory in Chile. Apart from entries in stellar catalogs, this binary is not present in the literature. Herein we present a full model of this binary star based on photometry and spectroscopy that has been derived using a range of custom and publicly available software packages. 

\begin{figure*}[htb!]
\includegraphics[width=\textwidth]{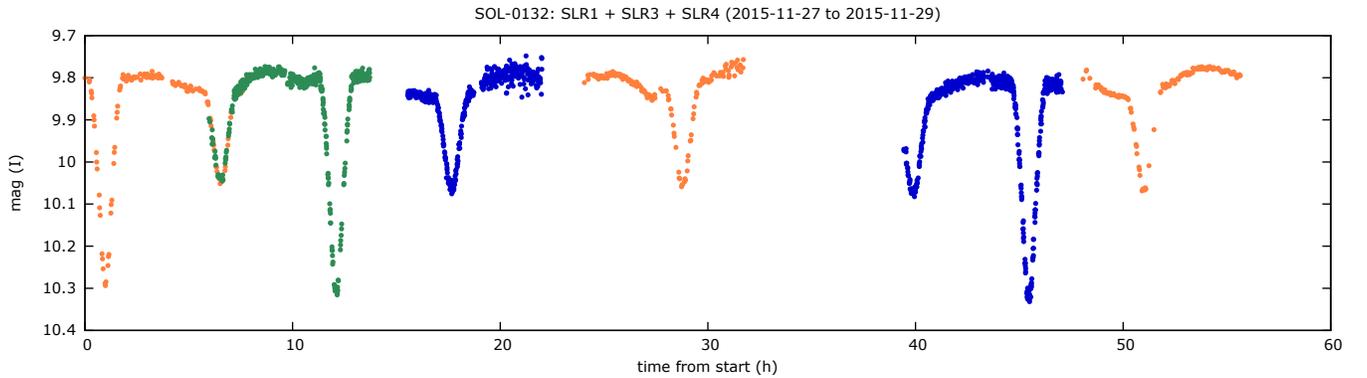}
\caption{J024946 55-h observing campaign with the Solaris telescope network: Solaris-1 (orange), Solaris-3 (blue) and Solaris-4 (green), I band - raw output from AiJ.}
\label{fig:J024946_55h}
\end{figure*}

\textbf{Photometry.} Photometric data has been collected during a long $\sim$55h run between November 27th and 29th, 2015 with the Solaris telescopes demonstrating the power of a global telescope network. Two gaps that have been caused by unfavorable weather conditions are present in the data set. The end of November is the time where the gap in the network coverage reaches its peak value (Fig. \ref{fig:NetworkCoverage}), hence it is not the optimal time to conduct campaigns requiring continuous coverage. Despite this, however,  we have been able to observe three primary and four secondary eclipses in just 2.3 days, an equivalent of $\sim$5 full orbital periods of the system. Altogether, multicolor photometry includes $\sim$3500 observations in V and I bands. A sample of I-band data reduced with AiJ is shown in Fig. \ref{fig:J024946_55h}. For modeling purposes I and V band data has been reduced using our custom pipeline and translated so that the levels agree with SLR3.

 \textbf{Spectroscopy}. 25 echelle spectra have been acquired between September 26th and October 22nd, 2015. Data has been reduced using \textsc{iraf}. RVs were computed using two methods: fitting a Gaussian profile to H-alpha emission for both stars and one-dimensional cross-correlation for the primary component (absorption lines from the secondary component are practically invisible in the spectra). An initial orbital fit was obtained using our custom \textsc{v2fit} code \citep{Konacki2010} using RVs obtained with both methods. This allowed us to verify that both methods lead to agreeing RV values. Errors of individual measurements were increased in quadrature during the fitting process to get $\chi^2\approx1$, a procedure that guarantees that values of parameters' errors are not underestimated. Then, photometric data along with RVs and initial orbital parameters were loaded into \textsc{phoebe}.

 \textbf{Modeling.} \textsc{phoebe} allows one to simultaneously fit photometric and spectroscopic data and, among many features, can also work with stellar spots. The physics behind the stellar spot functionality is rather simple -- the surface of the star that features a spot has lower (or higher) emergent intensity than spot-less parts of the star. The shape of J024946's light curve immediately reveals that stellar spots are present on at least one star. Moreover, the system shows activity in the X-ray domain and emission in the spectral region of H-$\alpha$ line, confirming that stellar spots are indeed very likely to be present. Following this logic, we have derived a model of J024946 that mimics the stellar spot with one large spot on the primary component.

\begin{figure}[htb!]
\centering
\includegraphics[width=\columnwidth,clip=true]{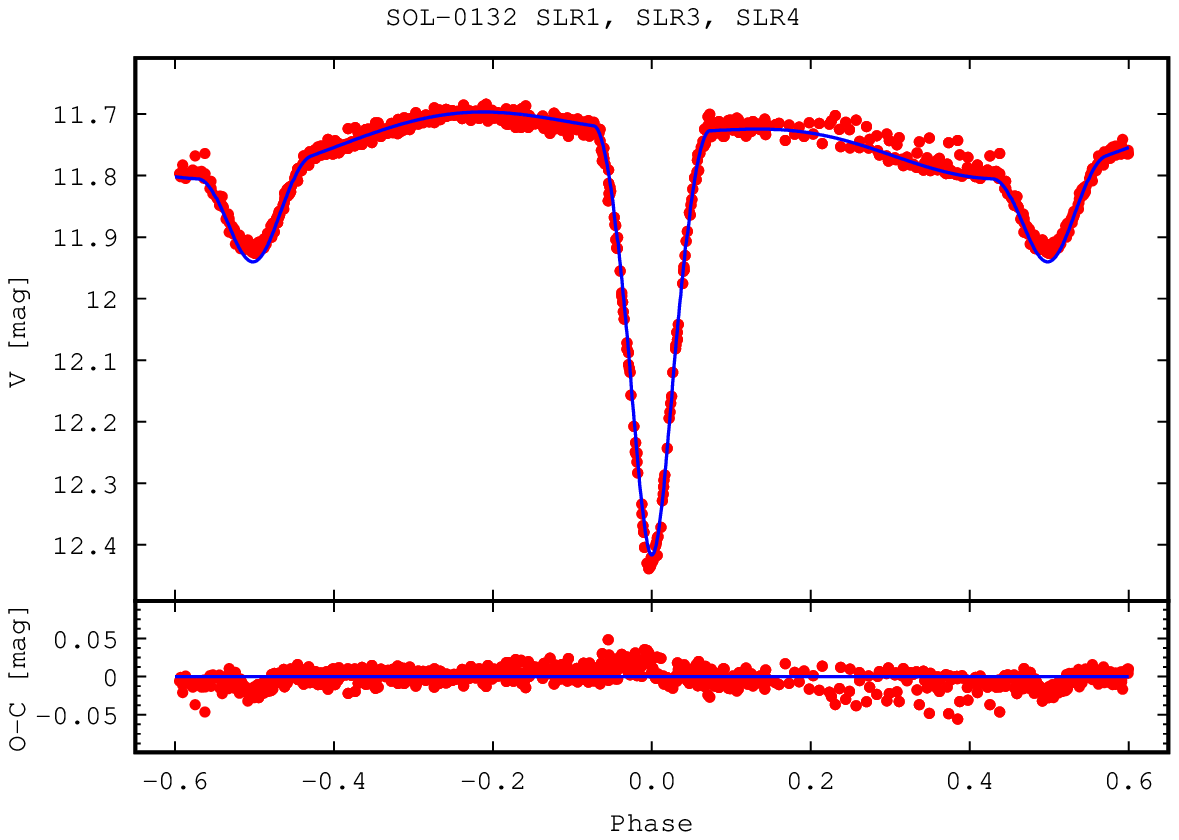}
\includegraphics[width=\columnwidth]{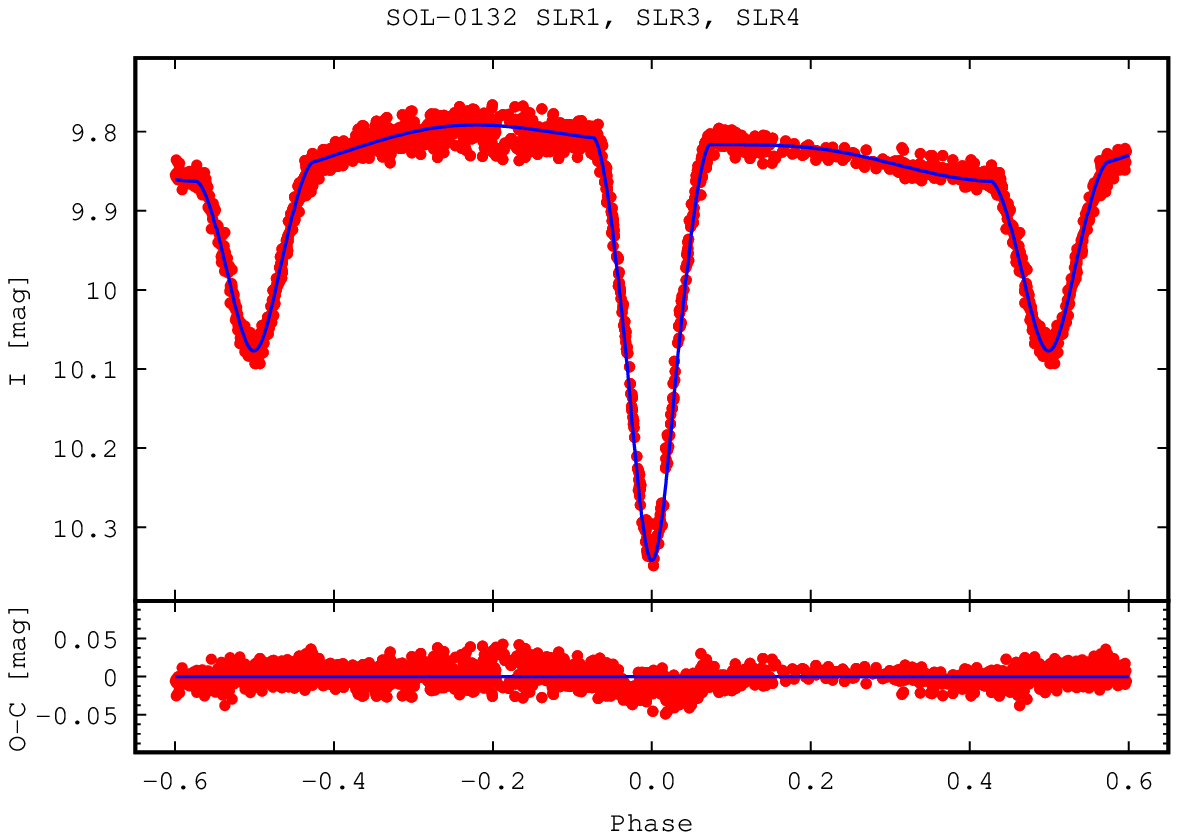}
\includegraphics[width=\columnwidth]{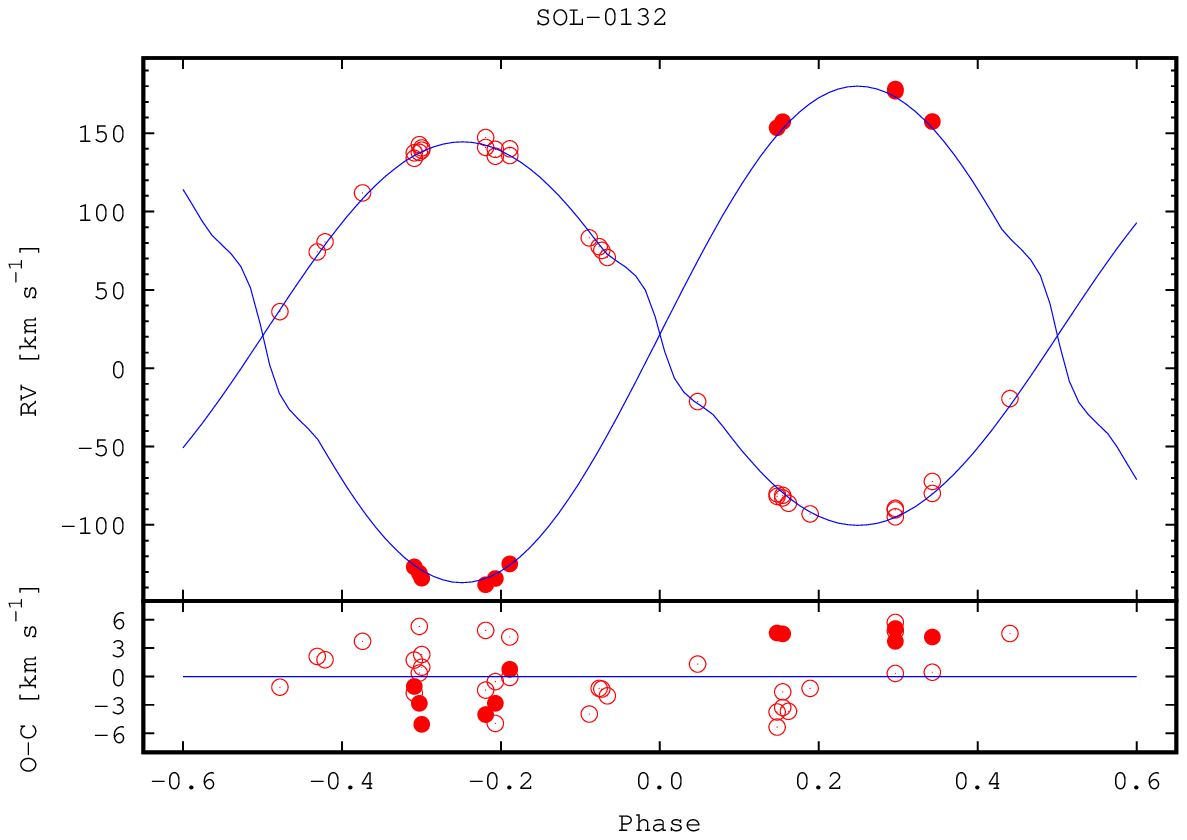}
\caption{J024946-3825.6 photometric (V and I bands) and spectroscopic data - measurements, models and residuals.}
\label{fig:J024946_models}
\end{figure}

Since the influence of a secondary component on the spectra is barely noticeable we have not succeeded in spectral disentangling and applied spectral analysis for co-added spectra to obtain atmospheric parameters of primary component. For this purpose, we used the software package \textit{Spectroscopy Made Easy} \citep[hereafter SME,][]{val96, val98}. Since line blending becomes more severe in the blue, what makes continuum placement and thus derived parameters less accurate \citep{val05}, we analyzed only 6 orders of the co-added spectrum which cover the wavelength region from 5992 to 6358 \AA. The list obtained from the Vienna Atomic Line Database \citep[VALD,][]{pis95, kup99} was used as atomic line data with the initial values described by \citet{val05} and \citet{kur92} model atmospheres. We used the value of $\log g$ derived from \textsc{phoebe} analysis and fixed it during spectral analysis, fitting $T_{\mathrm{eff}}$, $[M/H]$ and $v\sin i$ for every spectral order. Final values of $T_{\mathrm{eff}}$, $[M/H]$ were calculated as a mean value of the results from separate orders. The variance between orders was treated as the uncertainty of every parameter. The value of $T_{\mathrm{eff1}}$ obtained from spectral analysis was then fixed in \textsc{phoebe}, which was used in order to calculate $T_{\mathrm{eff2}}$. The resulting models are shown in Fig. \ref{fig:J024946_models}.

Absolute values of stellar and orbital parameters with its uncertainties were calculated with \textsc{jktabsdim} \citep{sou04a, sou04b} and are presented in Tab. \ref{tab:ModelsSolutions}. Since the value of \textit{v sin i} from spectral analysis was consistent with the value of rotational velocity calculated under the assumption of tidal locking $v_{syn}$ we adopted the latter value as a final one.

\begin{figure}[htb!]
\centering
\includegraphics[angle=-90,width=\columnwidth]{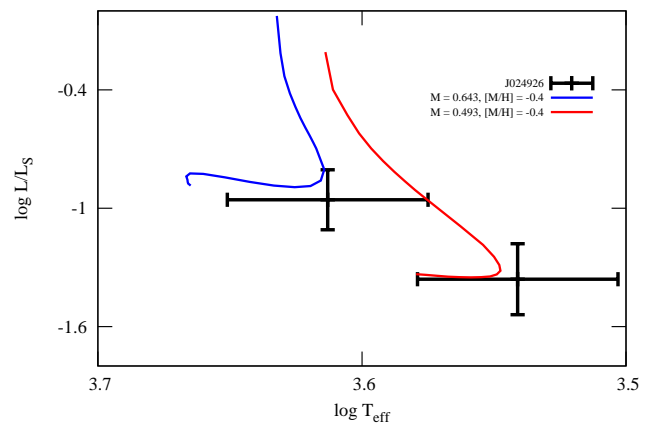}
\caption{YY evolutionary tracks for components of J024946-3825.6. Blue color represents a track calculated for primary component mass and system metallicity $[M/H]$ = -0.4, while red - secondaryÕs track.}
\label{fig:J024946_tracks}
\end{figure}

\begin{figure}[htb!]
\centering
\includegraphics[width=\columnwidth]{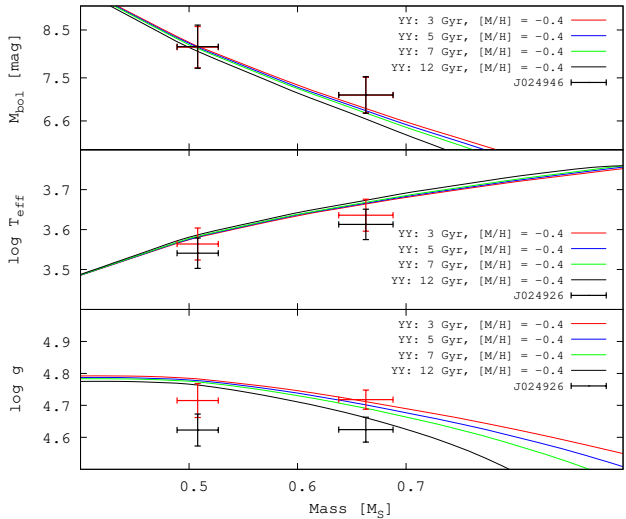}
\caption{YY isochrones for components of J024946-3825.6 calculated assuming system metallicity of $[M/H]$ = -0.4. Black symbols represent initial values of $log g$ and $T_{eff}$, red - recalculated after decreasing stellar radii (see text).}
\label{fig:J024946_iso}
\end{figure}

Yonsei-Yale \citep[hereafter YY, ][]{yi01} tracks and isochrones were used in order to check evolutionary status and age of the system. Evolutionary tracks calculated for given masses and the value of metallicity derived from spectral analysis ($[M/H] = -0.4$) are presented in Fig. \ref{fig:J024946_tracks} and imply that J024946-3825.6 is a main-sequence system.  Age estimation was based on fitting the isochrones of given metallicity to the observational data using three relationships: $M_\mathrm{bol}$ - mass, $\log T_\mathrm{eff}$ - mass, $\log g$ - mass and is presented in Fig. \ref{fig:J024946_iso}. As can be seen from the presented plots, stellar evolution models fail to reproduce properties like temperature and radius (thus $\log g$) obtained from our analysis. It is related to the known issue of low-mass stars discrepancy with stellar evolution models \citep[e.g.,][]{chab07,mor09, hel11} - objects are larger and cooler than expected. Studies \citep{fei12} show that model radii under-predict observed values up to a dozen percent. However model calculations are in good agreement with the observational mass - luminosity (thus $M_\mathrm{bol}$) relationship. Assuming that studied stars are 10$\%$ larger than model predictions, we re-calculated stellar parameters for the values of radius of $R_1-10\%$ and $R_2-10\%$ and corresponding effective temperatures (keeping constant value of luminosity) and presented it in Fig. \ref{fig:J024946_iso} using red color. It is clearly seen that the new values reproduce stellar models and the system age is $\sim$ 7 Gyr. 
  
\begin{figure}[htb!]
\centering
\includegraphics[width=\columnwidth]{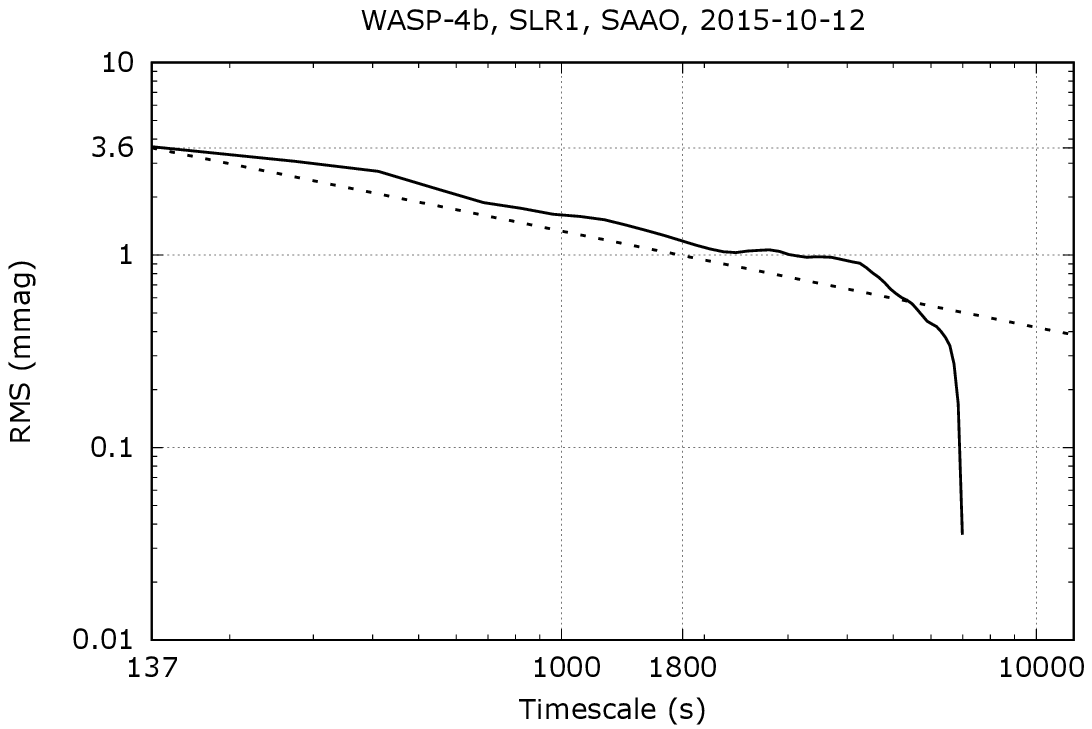}
\includegraphics[width=\columnwidth]{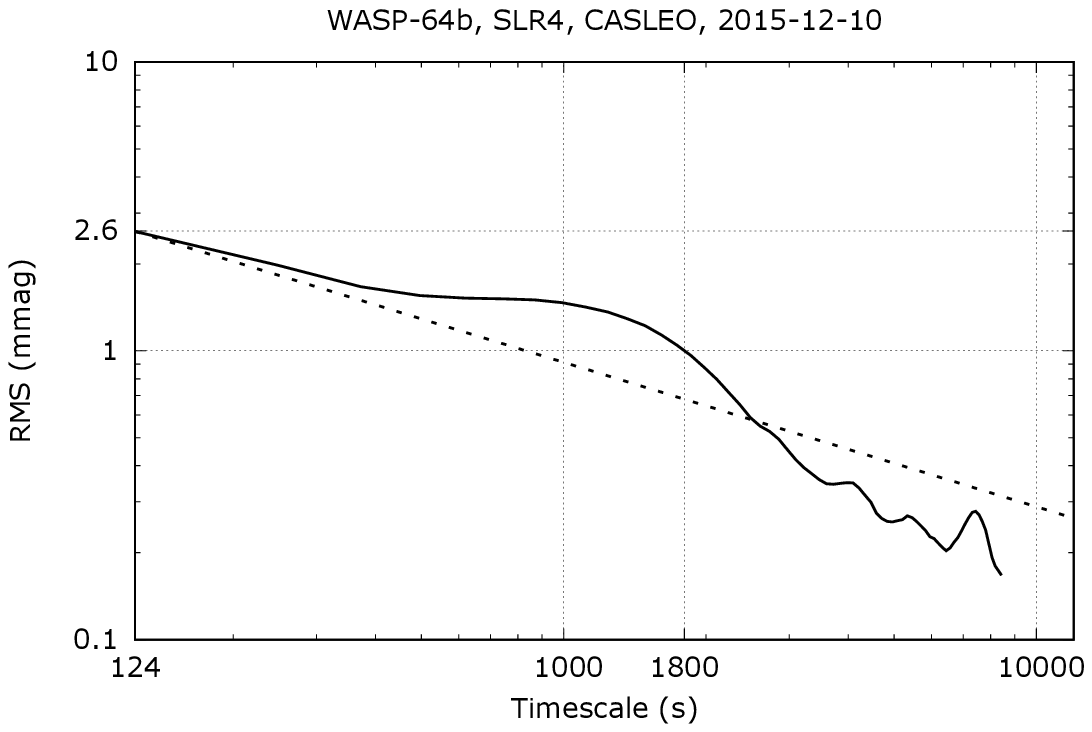}
\includegraphics[width=\columnwidth]{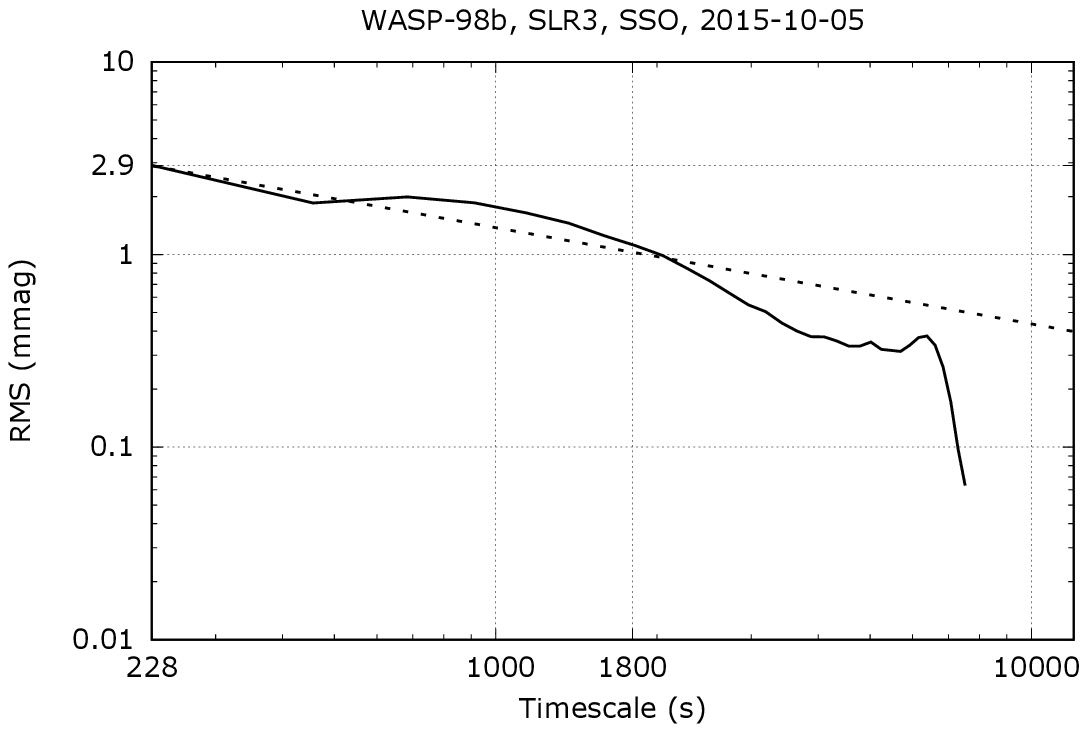}
\caption{Allan variance plots of transit residuals.}
\label{fig:transits-avar}
\end{figure}

\begin{figure}[htb!]
\centering
\includegraphics[width=\columnwidth]{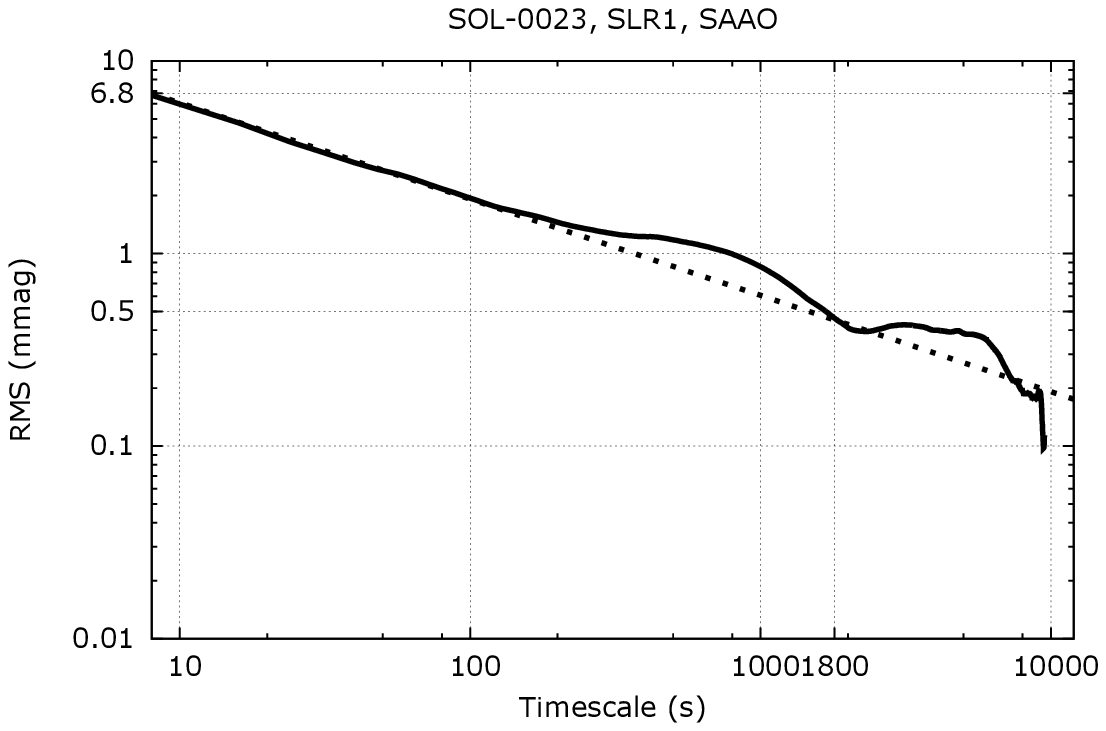}
\caption{Allan variance plots of SOL-0023, one of the timing targets.}
\label{fig:sol-0023-avar}
\end{figure}

\subsubsection{Photometric precision} 
In the previous sections we have demonstrated concrete evidence that proves the capability of the Solaris Network to acquire high quality data that can be used to tangle astrophysical problems of that require photometric measurements. Exoplanetary transits reduced with AiJ served as an initial test of the individual capabilities of the telescopes. Using off-the-shelf software allowed us to eliminate potential data reduction errors and prepare a benchmark for testing a dedicated pipeline. We follow an approach that is  similar to what has bee presented in \cite{Swift2015}, where the Minerva system's capabilities are benchmarked with the help of transits. The remaining targets were used to further demonstrate the precision achievable with the network. One of the best ways to quantify the photometric quality is to analyze the scatter of the O-C values that are a product of model fitting. This has been done for the following systems: Wasp-4b, Wasp-64b, Wasp-98b, SOL-0132 (V and I bands, three sites), PG-1338 and SOL-0023. The results are presented in Tab. \ref{tab:precision} and represent a wide range of cadence values from 8 to 500 s. In case of transits the fit RMS' values are comparable with the formal photometric errors that were computed during the data reduction process. Similarly, SOL-0023 shows good accordance between these values. It is worth noting that the formal errors themselves have small scatter. SOL-0023's data has been reduced using our dedicated pipeline. In case of SOL-0132 and and PG-1336 the average formal errors are smaller than the fit's RMS values. This has its origin in the modeling process. In both cases astrophysical models are computed and these have their limitations. In this case, however, the obtained astrophysical parameters and their formal errors serve as the measure of the data quality. It is important to measure and understand the sources of photometric errors but the end product is the key factor in quantifying the strength of the hardware and software medley. The precision in mass determination is 1.6 and 1.7 \% for PG-1336, 3.8 and  3.7 \% for SOL-0132. Radial velocity measurements' quality is the main factor, but for stellar radii, it is photometry that plays the crucial role. We obtain 1.4 \% precision for radii of the PG1336 components 4.5 and 5.2\% for SOL-0132. The latter is a very interesting case and has been described in detail in the previous section.

Allan variance has been computed for transits and SOL-0023. The plots are presented in Figs \ref{fig:transits-avar}-\ref{fig:sol-0023-avar}. White noise equivalents with appropriate standard deviation values are over-plotted to show the main components of the O-C scatter. In case of transits, 1 mmag precision is achievable in 30-minute timescales, in case of SOL-0023 this timescale is equivalent to 0.5 mmag precision. The last case is particularly interesting in terms of atmospheric scintillation due to the short 2.5-6 s exposure times. Based on the formula provided by \cite{Osborn2015}, the scintillation component accounts for 4.7 to 3.0 mmag of the final photometric residuals. 
The goal of the project is to precisely characterize a limited number of eclipsing binaries, predominately for eclipse timing. The pool of targets is ca. 300 binaries. During most nights no more than 1-2 targets (i.e. fields) per night are scheduled on a particular telescope. For this reasons, the performance of the network telescopes can be best tested on specific, know from literature targets that are typically characterized with high cadence (or relatively high), high precision observations (transits, pulsations, eclipsing binaries). The nature of photometric data reduction from systems such as Solaris (narrow field of view, 13-21 arcmin) is very specific and significantly different from wide-angle survey-type systems. Targets have been carefully chosen so that comparison stars are present in the field. Very often, target fields required an offset to avoid bright stars being saturated and at the same time move good comparison stars into the field. Hence the need for online astrometry and constant monitoring of the field. Solaris telescopes are high-end optical systems that produce flat fields, PSFs are well sampled with a high-grade CCD camera, very low read noise and negligible dark current. Exposure times for most targets need to updated online based on seeing conditions to react to changes in the observing conditions, keep the pixel counts in the desired range and avoid overexposure of field stars. These intrinsic properties of the Solaris systems influence the way data is handled and also scheduled. Wide-field systems usually have an abundance of comparison stars that are evenly spread across the field. This is not the case for Solaris. The final photometric errors strongly depend on the field and exposure time and, as demonstrated, are in the low mmag range and are consistent with modeling results.

\begin{deluxetable}{lcccc}
\tablewidth{0pt}
\tabletypesize{\scriptsize}								
\tablecaption{Formal photometric errors and fit RMS values for targets with computed models.}									
\tablehead{\colhead{Object} &\colhead{SLR} & \colhead{Cadence} & \colhead{Photometric error} & \colhead{Fit RMS} \\
				            &			     & \colhead{(s)}	     & \colhead{avg/min/max (mmag)}					  & \colhead{(mmag)}}								
\startdata									
Wasp-4b  		&1	& 137 		& 2.9 / 2.8 / 3.0  & 3.6  \\
Wasp-64b		&4	& 124 		& 2.1 / 1.7 / 3.1  & 2.6  \\
Wasp-98b		&3	& 229 		& 3.5 / 3.4 / 3.6  & 2.9  \\
SOL-0132 (V) 	&1	& 500		& 1.9 / 1.6 / 3.6  &        \\
			&3	& 46 - 56		& 5.7 / 4.3 / 8.6  & 8.7	   \\
			&4	& 140 - 155	& 2.6 / 1.7 / 3.3  &	    \\
SOL-0132 (I) 	&1	& 85 - 150		& 3.1 / 2.6 / 7.4  &        \\
			&3	& 26 - 31		& 10.0 / 7.4 / 18.0  & 10	   \\
			&4	& 140 - 155	& 4.9 / 3.3 / 8.2  &	    \\
PG-1336 		& 2	& 25			& 2.8 / 2.4 / 5.0 &  7.2 \\
SOL-0023		& 1	& 8			& 5.4 / 3.7 / 7.9 &  6.8
\enddata									
\label{tab:precision}									
\end{deluxetable}

%

\section{Summary}
\label{sec:Summary}

In this paper we have presented a global network of autonomous telescopes called Solaris. We have described the motivation that pushed us towards designing and building this scientific infrastructure consisting of four observatories located in the Southern Hemisphere. A thorough description of the network as a whole has been provided along with many technical details of the individual observatory components. Our design approach assumed the use of off-the-shelf components whenever possible, industrial standards and technologies wherever practical. Even the best design with carefully selected hardware will be useless without proper integration. In this paper we have described our software engineering approach, assumptions and prerequisites that led us to designing and building a custom observatory control software suite that allows the network to operate efficiently with minimal human intervention. Our modern approach and the use of industry approved computer science technologies allow the network to operate in such a way that reliability, availability and absolute efficiency metrics are in the high 90\% range. The autonomy of the system starts with observation scheduling and job distribution, through observatory control, image acquisition and ends with data transfer to the project's headquarters. Thanks to a small but dedicated team of professionals with profound experience in astrophysics, computer science and engineering we have reached the expected goals of the project. We have selected a set of eight targets that were used for scientific commissioning of the network. The goal was to verify that certain observing tasks can be accomplished with the network. Individual observing capabilities were verified during exoplanetary transit observations. Data has been reduced and analyzed with available software packages and our own software. Fitting transit models revealed residuals RMS values between 2.9 and 3.6 mmag. These have been compared with results available in the literature. We have also investigated three objects that are taking part in the timing campaign -- KZ Hya, RR Cae and SOL-0023. For these we have computed initial models and confirmed that formal photometric errors agree with RMS scatter. The commissioning campaign also included an interesting eclipsing sdB system. This particular target was used as the testbed for photometric precision and cadence. Again, using available software tools we were able to identify and distinguish pulsation semiamplitudes down to 2.1 mmag and construct a model of the system with the help of RV data available in the literature. Finally, we have presented a full detached eclipsing binary model based on photometric data obtained with three telescopes observing the same target in a continuous mode. Thanks to the global network we were able to cover more than four cycles in 55 hours with small gaps. RV measurements from the Chiron spectrograph supplement photometric data. We conclude that the commissioning results satisfy the project's assumptions.

Apart from showing the scientific potential of the network we have also described lessons learned from building and operating the network, major faults and problems that had to be resolved. 

Instrumental projects, such as Solaris, are mainly engineering undertakings, especially during the design and construction phases. Although most components of the observatories are off-the-shelf devices, their integration requires significant effort that builds expertise. If properly managed and financed, this know-how can gain commercial value. Several products that have been developed for the project have reached a level of maturity that allowed them to be commercialized via spin-out companies that were established by the project's team members. These products include the Abot software suite\footnote{http://sybillatechnologies.com} that runs the entire network, 2$\pi$Sky\footnote{http://www.cilium.pl}, the embedded cloud monitoring system, ObservatoryWatch, the PLC supervision system and several smaller software and hardware components that have been found to have market value as parts of larger systems. 

We believe that this is the proper way of running scientific instrumental projects, even in astrophysics, that initially seem not to have commercializable value. In fact, this approach brings benefit to the scientific part of the undertaking when financing of the project ceases. When this happens, further development is still possible thanks to private funding and cooperation with industry partners that initially derived from the scientific project. 

\acknowledgments

We are grateful to the technical and administration staff in SAAO, SSO and CASLEO for their help during the construction and commissioning of the telescopes.
This work is supported by the European Research Council through a Starting Grant, the National Science Centre through grant 5813/B/H03/2011/40, the Polish Ministry of Science and Higher Education through grant 2072/7.PR/2011/2 and the Foundation for Polish Science through \textit{Idee dla Polski} funding scheme. K.G.H. acknowledges support provided by the National Science Center through grant 2016/21/B/ST9/01613. P.S. acknowledges support provided by the National Science Center through grant 2011/03/N/ST9/03192. M.R. acknowledges support provided by the National Science Center through grant 2011/01/N/ST9/02209.

\bibliography{Solaris_PASP}

\appendix

Radial velocity measurements.

\clearpage
\begin{deluxetable*}{crrrrrrr}
\tablecaption{RV measurements for J024946 }
\tablehead{\colhead{}		&	\colhead{MJD}	&	\colhead{RV$_1$}	&	\colhead{$\sigma$RV$_1$}	&	\colhead{O-C}	&	\colhead{RV$_2$}	&	\colhead{$\sigma$RV$_2$}	&	\colhead{O-C}}
\startdata
A024946	&	57291.131084	&	140.55187	&	3.64005	&	 1.51214	&	-133.98832	&	1.56205	&	-1.10130	\\	
H-alpha	&	57291.182418	&	140.01709	&	3.64005	&	 5.89404	&	-124.93019	&	1.56205	&	 1.37538	\\	
	&	57292.267897	&	-81.09564	&	3.64005	&	-2.53840	&	 157.35691	&	1.56205	&	-1.51432	\\	
	&	57292.333549	&	-89.46779	&	3.64005	&	 2.97774	&	 176.84990	&	1.56205	&	-0.67672	\\	
	&	57292.333549	&	-90.38141	&	3.64005	&	 2.06412	&	 178.22032	&	1.56205	&	 0.69370	\\	
	&	57293.281613	&	-72.29459	&	3.64005	&	 4.61514	&	 157.47865	&	1.56205	&	 0.82015	\\	
	&	57302.246941	&	137.82095	&	3.64005	&	 0.23919	&	-130.78077	&	1.56205	&	 0.15465	\\	
	&	57302.285603	&	140.92891	&	3.64005	&	-0.80331	&	-138.17934	&	1.56205	&	-1.68838	\\	
	&	57303.170495	&	134.00611	&	3.64005	&	-1.88401	&	-126.82991	&	1.56205	&	 1.84111	\\	
	&	57303.217776	&	135.27811	&	3.64005	&	-4.31503	&	-134.23722	&	1.56205	&	-0.60944	\\	
	&	57304.308438	&	-81.77780	&	3.64005	&	-7.74631	&	 153.47711	&	1.56205	&	 0.68421	\\	
	&	&	&	&	&	&	&								\\	
A024946	&	57291.131084	&	 139.21468	&	2.25853	&	-1.15177	&		&		&		\\	
1DCCF	&	57291.182418	&	 135.75045	&	2.26011	&	-0.01644	&		&		&		\\	
	&	57292.165721	&	  70.71910	&	2.25680	&	 1.27476	&		&		&		\\	
	&	57292.218362	&	 -21.31230	&	2.25570	&	-3.95667	&		&		&		\\	
	&	57292.267897	&	 -82.74540	&	2.25540	&	-0.32894	&		&		&		\\	
	&	57292.333549	&	 -94.87022	&	2.25834	&	 2.12739	&		&		&		\\	
	&	57293.197730	&	 -86.17637	&	2.25588	&	-0.72970	&		&		&		\\	
	&	57293.210309	&	 -92.98616	&	2.25662	&	 1.67862	&		&		&		\\	
	&	57293.281613	&	 -79.89517	&	2.26097	&	 1.14625	&		&		&		\\	
	&	57293.364496	&	  36.07290	&	2.25767	&	-4.51721	&		&		&		\\	
	&	57302.127963	&	 -19.44835	&	2.26013	&	 2.52565	&		&		&		\\	
	&	57302.187515	&	  74.19036	&	2.25847	&	-1.41285	&		&		&		\\	
	&	57302.246941	&	 142.74726	&	2.25807	&	 3.09871	&		&		&		\\	
	&	57302.285603	&	 147.24704	&	2.26013	&	 4.30922	&		&		&		\\	
	&	57302.346052	&	  83.15756	&	2.26656	&	-2.08771	&		&		&		\\	
	&	57303.118465	&	  80.61606	&	2.29825	&	-1.72438	&		&		&		\\	
	&	57303.170495	&	 137.49415	&	2.25767	&	-0.56366	&		&		&		\\	
	&	57303.217776	&	 139.69817	&	2.25531	&	-0.84690	&		&		&		\\	
	&	57303.278211	&	  77.55610	&	2.25700	&	 0.76836	&		&		&		\\	
	&	57304.308438	&	 -80.17848	&	2.25400	&	-0.80282	&		&		&		\\	
	&	57315.184084	&	 111.93272	&	2.25640	&	 0.54823	&		&		&		\\	
	&	57317.176350	&	  75.19947	&	2.25448	&	 0.74512	&		&		&		\\	
\enddata
\end{deluxetable*}

\end{document}